\documentclass{article}

\usepackage{arxiv}

\usepackage[utf8]{inputenc} % allow utf-8 input
\usepackage[T1]{fontenc}    % use 8-bit T1 fonts
\usepackage{hyperref}       % hyperlinks
\usepackage{url}            % simple URL typesetting
\usepackage{booktabs}       % professional-quality tables
\usepackage{amsfonts}       % blackboard math symbols
\usepackage{nicefrac}       % compact symbols for 1/2, etc.
\usepackage{microtype}      % microtypography
\usepackage{lipsum}		% Can be removed after putting your text content
\usepackage{graphicx}
\usepackage[numbers]{natbib}
\usepackage{doi}
\usepackage{fontenc, inputenc, calc, indentfirst, fancyhdr, graphicx, epstopdf, lastpage, ifthen, lineno, float, amsmath, setspace, enumitem, mathpazo, booktabs, titlesec, etoolbox, tabto, xcolor, soul, multirow, microtype, tikz, totcount, changepage, paracol, attrib, upgreek, cleveref, amsthm, hyphenat, natbib, hyperref, footmisc, url, geometry, newfloat, caption}
\usepackage[utf8]{inputenc} % allow utf-8 input
\usepackage[T1]{fontenc}    % use 8-bit T1 fonts
\usepackage{hyperref}       % hyperlinks
\usepackage{url}            % simple URL typesetting
\usepackage{booktabs}       % professional-quality tables
\usepackage{amsfonts}       % blackboard math symbols
\usepackage{nicefrac}       % compact symbols for 1/2, etc.
\usepackage{microtype}      % microtypography
\usepackage{amsmath,amsthm,amssymb}
\usepackage{mathtools}
\usepackage{algorithmic}
\usepackage{algorithm}
\usepackage{booktabs}
\usepackage{graphicx} 
\usepackage{caption}
\usepackage{subcaption}

\DeclareMathOperator*{\argmax}{arg\,max}
\DeclareMathOperator*{\argmin}{arg\,min}

\newcommand{\mmd}{\text{\sc{mmd}}}
\newcommand{\mmdhat}{\widehat{\text{\sc{mmd}}^2}}

\definecolor{redtol}{RGB}{187,85,102}
\definecolor{bluetol}{RGB}{0, 68, 136}
\title{Understanding collections of related datasets using dependent MMD coresets}

\author{Sinead A.\ Williamson\\Department of Statistics and Data Science\\ University of Texas at Austin \\ \texttt{sinead@austin.utexas.edu}
	\And
	Jette Henderson \\
	CognitiveScale \\
	\texttt{jhenderson@cognitivescale.com} \\
}
\date{}

\renewcommand{\shorttitle}{Understanding collections of related datasets using dependent MMD coresets}

\hypersetup{
pdftitle={AUnderstanding collections of related datasets using dependent MMD coresets},
pdfsubject={},
pdfauthor={Sinead A. Williamson, Jette Henderson},
pdfkeywords={coresets, data summarization, maximum mean discrepency, interpretability},
}

\begin{document}
\maketitle
% Abstract (Do not insert blank lines, i.e. \\) 
\begin{abstract}
Understanding how two datasets differ can help us determine whether one dataset under-represents certain sub-populations, and provides insights into how well models will generalize across datasets. Representative points selected by a maximum mean discrepency (MMD) coreset can provide interpretable summaries of a single dataset, but are not easily compared across datasets. In this paper we introduce dependent MMD coresets, a data summarization method for collections of datasets that facilitates comparison of distributions. We show that dependent MMD coresets are useful for understanding multiple related datasets and understanding model generalization between such datasets.
\end{abstract}

\section{Introduction}

When working with large datasets, it is important to understand your data.  If a dataset is not representative of your population of interest, and no appropriate correction is made, then models trained on this data may perform poorly in the wild. Sub-populations that are underrepresented in the training data are likely to be poorly served by the resulting algorithm, leading to unanticipated or unfair outcomes---something that has been observed in numerous scenarios including medical diagnoses \citep{LarNiePetMilFer2020,CheJohSon2018} and image classification \citep{BuoGeb2018,ShaHalYonAtwWilScu2017}. 

In low-dimensional settings, it is common to summarize data using summary statistics such as marginal moments or label frequencies, or to visualize univariate or bivariate marginal distributions using histograms or scatter plots. As the dimensionality of our data increase, such summaries and visualizations become unwieldy, and ignore higher-order correlation structure. In structured data such as images, such summary statistics can be hard to interpret, and can exclude important information about the  distribution \cite{alexander2014summary,lauer2018role}---the per-pixel mean and standard deviation of a collection of images tells us little about the overall distribution. Further, if our data isn't labeled, or is only partially labeled, we cannot make use of label frequencies to assess class balance.

In such settings, we can instead choose to present a set of exemplars that capture the diversity of the data. This is particularly helpful for structured, high-dimensional data such as images or text, that can easily be qualitatively assessed by a person. Random sampling provides a simple way of obtaining such a collection, but there is no guarantee that a given data point will be close to one of the resulting representative points---particularly in high dimensions or multi-modal distributions. A number of algorithms have been proposed to provide more representative exemplars \cite{KauRou1987,bien2011prototype,MakJos2017,MakJos2018,KimKhaKoy2016,wilson2000reduction,GurDhuCecAgg2019,CheWelSmo2010}. Many of these algorithms can be seen as constructing a coreset for the dataset---a (potentially weighted) set of exemplars that behave similarly to the full dataset under a certain class of functions. In particular, coresets that minimize the maximum mean discrepency \citep[MMD,][]{GreBorRasSchSmo2010} between coreset and data have recently been used for understanding data distributions  \cite{KimKhaKoy2016,GurDhuCecAgg2019}. Further, evaluating models on such MMD-coresets have been shown to aid in understanding model performance \citep{KimKhaKoy2016}.

In addition to summarizing a single dataset, we may also wish to compare and contrast multiple related datasets. For example, a company may be interested in characterizing differences and similarities between different markets. A machine learning practitioner may wish to know whether their dataset is similar to that used to train a given model. A researcher may be interested in understanding trends in posts or images on social media. Here, summary statistics offer interpretable comparisons: we can plot the mean and standard deviation of a given marginal quantity over time, and easily see how it changes \cite{PraTsc2003,HohWonKerPat2020}. By contrast, coresets are harder to compare, since the exemplars selected for two datasets $X_1$ and $X_2$ will not in general overlap.

In this paper, we introduce dependent MMD coresets $\{\mathbb{Q}_t\}_{t\in\mathcal{T}}$ for a collection $\{X_t\}_{t\in\mathcal{T}}$ of related datasets. Each marginal coreset $\mathbb{Q}_t$ is an MMD coreset for the corresponding dataset $X_t$---i.e., it provides a set of weighted exemplars that can be used to understand the dataset $X_t$ and evaluate model performance in the context of that dataset. However, the marginal coresets have shared support: a shared set of exemplars are used to describe all datasets, but the weight assigned to each exemplar varies between datasets.

This shared support makes it easy to understand differences between datasets. Consider comparing two datasets of faces. Independent MMD coresets would provide two disjoint set of weighted exemplars; visually assessing the similarity between these coresets involves considering both the similarities of the images \textit{and} the similarities in the weights. Conversely, with a dependent MMD coreset, the exemplars would be shared between the two datasets. Similarity can be assessed by considering the relative weights assigned in the two marginal coresets. This in turn leads to easy summarization of the difference between the two datasets, by identifying exemplars that are highly representative of one dataset, but less representative of the other.

We begin by considering existing coreset methods for data and model understanding in Section~\ref{sec:bg}, before discussing their limitations and proposing our dependent MMD coreset in Section~\ref{sec:method}. A greedy algorithm to select dependent MMD coresets is provided in Section~\ref{sec:alg}. In Section~\ref{sec:experiments} we evaluate the efficacy of this algorithm, and show how the resulting dependent coresets can be used for understanding collections of image datasets and probing the generalization behavior of machine learning algorithms.

\section{Background and related work}\label{sec:bg}

\begin{table}[]
    \centering
    \begin{tabular}{c|l}
        $\mathcal{T}$ & set that indexes datasets and associated measures  \\
        $X_t = (x_{t,1},\dots, x_{t,n_t}) \in \mathcal{X}^{n_t}$ & a dataset indexed by $t\in\mathcal{T}$\\
        $\mathbb{P}_t$ & true distribution at $t\in\mathcal{T}$, $X_t\sim\mathbb{P}_t$\\
        $U = (u_1, \dots, u_{n_u})$& set of candidate locations\\
        %$\mathbf{w}_t=(w_{t,1},\dots, w_{t, n_t})$ & a vector of weights such that $\sum_i w_{t, i} = 1$\\ 
        $\delta_u$ & Dirac measure (i.e.\ point mass) at $u$.\\
        $\mathbb{Q}_t$ & a probability measure used to approximate $\mathbf{P}_t$, that takes \\
        & the form $\sum_{i\in S} w_{t,i}\delta_{u_i}$, where $S\subset [n_u]$
    \end{tabular}
    \caption{Notation used in this paper}
    \label{tab:my_label}
\end{table}
\subsection{Coresets and measure-coresets}

A coreset is a ``small'' summary of a dataset $X$, which can act as a proxy for the dataset under a certain  class of functions $\mathcal{F}$. Concretely, a weighted set of points $\{(w_i, u_i)\}_{i\in S}$ are an $\epsilon$ strong coreset for a size-$n$ dataset $X$ with respect to $\mathcal{F}$ if
$$\left\lvert \frac{1}{n}\sum_{i=1}^{n}f(x_i) - \frac{1}{|S|}\sum_{j\in S}w_j f(u_j)\right\rvert \leq \epsilon$$
for all $f\in\mathcal{F}$ \cite{agarwal2004approximating}.

A measure coreset \cite{ClaSol2018} generalizes this idea to assume that $X$ are independently and identically distributed samples from some distribution $\mathbb{P}$. A measure $\mathbb{Q}$ is an $\epsilon$-measure coreset for $\mathbb{P}$ with respect to some class $F$ of functions if
\begin{equation}\sup_{f\in\mathcal{F}}\left\lvert \mathbb{E}_{X\sim \mathbb{P}}\right[f(X)\left]  - \mathbb{E}_{Y\sim \mathbb{Q}}\right[f(Y)\left]\right\rvert \leq \epsilon.\label{eqn:measure_coreset}\end{equation}
The left hand side of Equation~\ref{eqn:measure_coreset} describes an integral probability metric \cite{muller1997integral}, a class of distances between probability measures parametrized by some class $\mathcal{F}$ of functions. Different choices of $\mathcal{F}$ yield different distributions (Table~\ref{tab:ipm}). 

\begin{table}
    \centering
    \begin{tabular}{c c c}
          Distance &$\qquad$ &$\mathcal{F}$ \\
         \hline
         1-Wasserstein distance & $\qquad$ &$\{f: ||\nabla f||_1\leq 1\}$ \\
         Maximum mean discrepancy & $\qquad$ &$\{f: ||f||_\mathcal{H} \leq 1\}$ for some RKHS $\mathcal{H}$ \\
         Total variation  & $\qquad$ &$\{f: ||f||_\infty \leq 1\}$
    \end{tabular}
    \caption{Some examples of integral probability metrics}
    \label{tab:ipm}
\end{table}

\subsection{MMD-coresets}
In this paper, we consider the case where $\mathcal{F}$ is the class of all functions that can be represented in the unit ball of some reproducing kernel Hilbert space (RKHS)  $\mathcal{H}$---a very rich class of continuous functions on $\mathcal{X}$. This corresponds to a metric known as the maximum mean discrepancy \citep[MMD,][]{GreBorRasSchSmo2010},

\begin{equation}\mmd(\mathbb{P}, \mathbb{Q}) = \sup_{f\in\mathcal{H}}\left\lvert \mathbb{E}_{X\sim \mathbb{P}}\right[f(X)\left]  - \mathbb{E}_{Y\sim \mathbb{Q}}\right[f(Y)\left]\right\rvert.\label{eqn:mmd_def}\end{equation}

An RKHS can be defined in terms of a mapping $\Phi: \mathcal{X}\rightarrow \mathcal{H}$, which in turn specifies a kernel function $k(x,x') = \langle \Phi(x), \Phi(x')\rangle_\mathcal{H}$. A distribution $\mathbb{P}$ can be represented in this space in terms of its mean embedding, $\mu_P = \mathbb{E}_\mathbb{P}\left[\Phi(x)\right]$. The MMD between two distributions equivalently can be expressed in terms of their mean embeddings,
$\mmd(\mathbb{P}, \mathbb{Q})^2 = ||\mu_P - \mu_Q||_\mathcal{H}^2$,

An $\epsilon$-MMD coreset for a distribution $\mathbb{P}$ is a finite, atomic distribution $\mathbb{Q} = \sum_{i\in S} w_i \delta_{u_i}$ such that $\mmd(\mathbb{P}, \mathbb{Q})^2\leq \epsilon^2$. We will refer to the set $\{u_i\}_{i\in S}$ as the support of $\mathbb{Q}$, and refer to individual locations in the support of $\mathbb{Q}$ as exemplars.

In practice, we are unlikely to have access to $\mathbb{P}$ directly, but instead have samples $X:=(x_1, \dots, x_{n})\sim \mathbb{P}$. If $\mathbb{Q}=\sum_{i\in S} w_i \delta_{u_i}$, we can estimate $\mmd(\mathbb{P}, \mathbb{Q})^2$ as

$$\mmdhat\left(X, \mathbb{Q}\right) = \frac{1}{n^2}\sum_{i=1}^{n}\sum_{j=1}^{n}k(x_i, x_j) + \sum_{i\in S}\sum_{j\in S} w_i w_j k(u_i, u_j) - \frac{2}{n}\sum_{i=1}^{n}\sum_{j\in S} w_jk(x_i, u_j).$$

We therefore define an $\epsilon$-MMD coreset for a dataset $X$ as a finite, atomic distribution $\mathbb{Q}$ such that $\mmdhat(X, \mathbb{Q})\leq \epsilon^2$---or equivalently, whose mean embedding $\mu_Q$ in $\mathcal{H}$ is close to the empirical mean embedding $\hat{\mu}_X$ so that $||\mu_Q-\hat{\mu}_X||_\mathcal{H}^2\leq \epsilon^2$.

A number of algorithms have been proposed that correspond to finding $\epsilon$-MMD coresets, under certain restrictions on $\mathbb{Q}$.\footnote{While most of these algorithms do not explicitly use coreset terminology, the resulting set of samples, exemplars or prototypes meet the definition of an MMD coreset for some value of $\epsilon$.} The kernel herding algorithm \cite{CheWelSmo2010,bach2012equivalence,lacoste2015sequential} can be seen as finding an MMD coreset $\mathbb{Q}$ for a known distribution $\mathbb{P}$, with no restriction on the support of $\mathbb{Q}$. The greedy prototype-finding algorithm used by \cite{KimKhaKoy2016} can be seen as a version of kernel herding, where $\mathbb{P}$ is only observed via a set of samples $X$, and where the support of $\mathbb{Q}$ is restricted to be some subset of a collection of candidates $U$ (often chosen to be the data set $X$). Versions of this algorithm that assign weights to the atoms in $\mathbb{Q}$ are proposed in \cite{GurDhuCecAgg2019}. 

As in \cite{KimKhaKoy2016} and \cite{GurDhuCecAgg2019}, in this paper we require the support of our coreset to be a subset of some finite set of candidates $U$, indexed by $1,\dots, n_U$. In other words, our measure coresets will take the form $\mathbb{Q} = \sum_{i\in S} w_i \delta_{u_i}$, where $S\subset [n_U]$.

\subsection{Coresets for understanding datasets and models}\label{sec:bg_understanding}

The primary application of coresets is to create a compact representation of a large dataset, to allow for fast inference on downstream tasks (see \cite{feldman2020introduction} for a recent survey). However, such compact representations have also proved beneficial in interpretation of both models and datasets.

While humans are good at interpreting visual data \citep{potter2014detecting}, visualizing large quantities of data can become overwhelming due to the sheer quantity of information. Coresets can be used to filter such large datasets, while still retaining much of the relevant information.

The MMD-critic algorithm \cite{KimKhaKoy2016} uses a fixed-dimension, uniformly weighted MMD coreset, which they refer to as ``prototypes'', to summarize collections of images. \cite{GurDhuCecAgg2019} extends this to use a weighted MMD coreset, showing that weighted prototypes allow us to better model the data distribution, leading to more interpretable summaries. \cite{ZheOuLexPhi2017} show how coresets designed to approximate a point cloud under kernel density estimation can be used to represent spatial point processes such as spatial location of crimes.

Techniques such as coresets that produce representative points for a dataset can also be used to provide interpretations and explanations of the behavior of models on that dataset. Case-based reasoning approaches use representative points to describe ``typical'' behavior of a model ~\cite{kim2014bayesian,aamodt1994case,murdock2003assessing,KimKhaKoy2016}. Considering the model's output on such representative points can allow us to understand the model's behavior. 

Viewing the model's behavior on a collection of ``typical'' points in our dataset also allows us to get an idea of the overall model performance on our data. Evaluating a model on a coreset can give an idea of how we expect it to perform on the entire dataset, and can help identify failure modes or subsets of the data where the model performs poorly. 

\subsection{Criticising MMD coresets}\label{sec:bg_mmd_critic}

While MMD coresets are good at summarizing a distribution, since the coreset is much smaller than the original dataset, there are likely to be outliers in the data distribution that are not well explained by the coreset. The MMD-critic algorithm supplements the ``prototypes'' associated with the MMD coreset with a set of ``criticisms''---points that are poorly modeled by the coreset \cite{KimKhaKoy2016} .

Recall from Equation~\ref{eqn:mmd_eqn} that the MMD between two distributions $\mathbb{P}$ and $\mathbb{Q}$ corresponds to the maximum difference in the expected value on the two spaces, of a function that can be represented in the unit ball of a Hilbert space $\mathcal{H}$. 
The function $f$ that achieves this maximum is known as the witness function, and is given by
$$f(x^*) = \mathbb{E}_{X\sim\mathbb{P}}[k(x^*, X]] - \mathbb{E}_{Y\sim\mathbb{Q}}[k(x^*, Y]$$

When we only have access to $\mathbb{P}$ via a size-$n$ sample $X$, and where $\mathbb{Q} = \sum_{i\in S} w_i \delta_{u_i}$, we can approximate this as

$$\hat{f}(x^*) = \frac{1}{n}\sum_{i=1}^n k(x^*, X_i) - \sum_{j\in S}w_j k(x^*, u_j)$$ 

Criticisms of an MMD coreset for a data set $X$ are selected as the points in $X$  with the largest values of the witness function. \citet{KimKhaKoy2016} show that the combination of prototypes and criticisms allow us to visually understand large collections of images: the prototypes summarize the main structure of the dataset, while the criticisms allow us to represent the extrema of a distribution. Criticisms can also augment an MMD coreset in a case-based reasoning approach to model understanding, by allowing us to consider model behavior on both ``typical'' and ``atypical'' exemplars.

\subsection{Dependent and correlated random measures}
Dependent random measures \cite{maceachern1999dependent,quintana2020dependent} are distributions over collections of countable measures $\mathbb{P}_t = \sum_{i=1}^\infty w_{t, i}\delta_{u_{t, i}}$, indexed by some set $\mathcal{T}$, such that the marginal distribution at each $t\in\mathcal{T}$ is described by a specific distribution. In most cases, this marginal distribution is a Dirichlet process, meaning that the $\mathbb{P}_t$ are probability distributions. Most dependent random measures either keep the weights $w_{t, i}$ or the atom locations $u_{t, i}$ constant accross $t$, to assist identifiability and interpretability.

In a Bayesian framework, dependent Dirichlet processes are often used as a prior for time-dependent mixture models. In settings where the atom locations (i.e.\ mixture components) are fixed but the weights vary, the posterior mixture components can be used to visualize and understand data drift \cite{de2004anova,dubey2013nonparametric}. The dependent coresets presented in this paper can be seen as deterministic, finite-dimensional analogues of these posterior dependent random measures.

\section{Understanding multiple datasets using coresets}\label{sec:method}
As we have seen, coresets provide a way of summarizing a single distribution. In this section, we discuss interpretational limitations that arise when we attempt to use coresets to summarize multiple related datasets (Section~\ref{sec:limitations}), before proposing dependent MMD coresets in Section~\ref{sec:dependent_mmd_coresets} and discussing their uses in Section~\ref{sec:method_understanding}. 

\subsection{Understanding multiple datasets using MMD-coresets}\label{sec:limitations}

If we have a collection $\{X_t\}_{t\in \mathcal{T}}$ of datasets, we might wish to find $\epsilon$-MMD coresets for each of the $X_t$, in the hope of not just summarizing the individual datasets, but also of easily comparing them. However, if we want to understand the relationships between the datasets, in addition to their marginal distributions, comparing such coresets in an interpretable manner is challenging.  

An MMD coreset selects a set $\{u_i\}_{i\in S}$ of points from some set of candidates $U$. Even if two datasets $X$ and $Y$ are sampled from the same underlying distribution (i.e., $X, Y\stackrel{\text{\small{iid}}}{\sim} \mathbb{P}$), and the set $U$ of available candidates is shared,  the optimal MMD-coreset for the two datasets will differ in general. Sampling error between the two distributions means that $\mmdhat(X, \mathbb{Q}) \neq \mmdhat(Y, \mathbb{Q})$ for any candidate coreset $Q$ unless $X\equiv Y$, and so the optimal coreset will typically differ between the two datasets.

\begin{figure}[h]
     \centering
     \begin{subfigure}{0.4\textwidth}
         \centering
         \includegraphics[width=\textwidth]{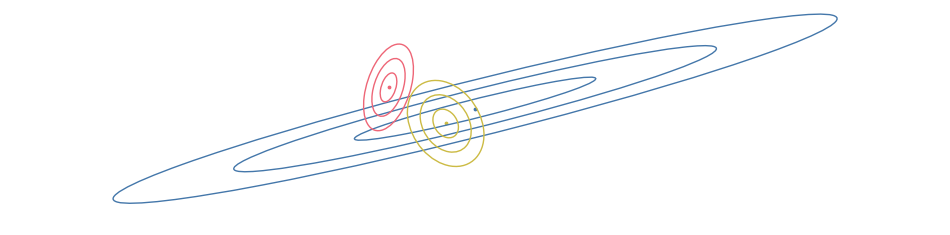}
         \caption{A mixture of three equally weighted Gaussians.}
         \label{fig:g3dist}
     \end{subfigure}
     \\
     \vspace{.1in}
     \begin{subfigure}{0.4\textwidth}
         \centering
         \includegraphics[width=\textwidth]{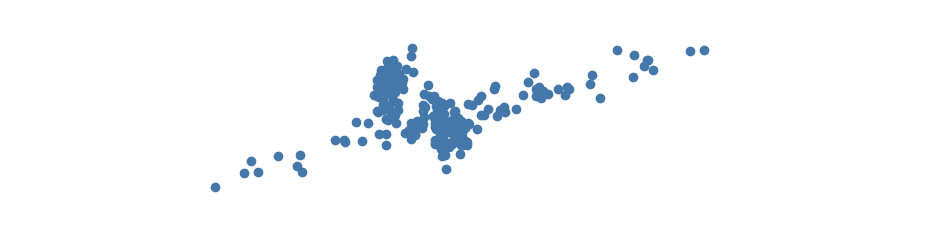}
         \caption{First sample of 250 observations from the distribution in Figure~\ref{fig:g3dist}.}
         \label{fig:g3s1}
     \end{subfigure}
     ~~~
     \begin{subfigure}{0.4\textwidth}
         \centering
         \includegraphics[width=\textwidth]{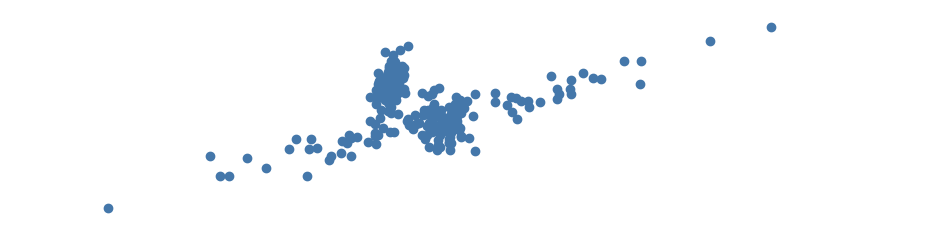}
     \caption{Second sample of 250 observations from the distribution in Figure~\ref{fig:g3dist}.}
         \label{fig:g3s2}
     \end{subfigure}\\
     \vspace{.1in}
     \begin{subfigure}{0.4\textwidth}
         \centering
         \includegraphics[width=\textwidth]{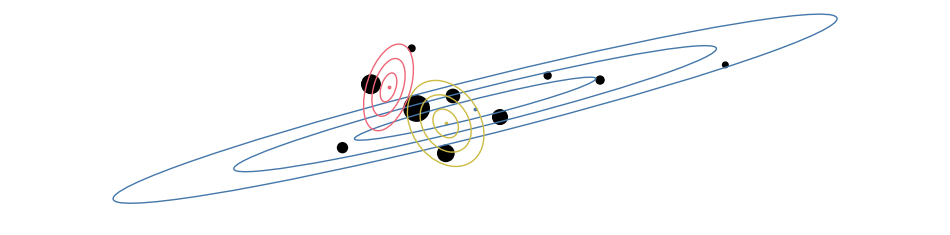}
     \caption{MMD coreset for the sample in Figure~\ref{fig:g3s1}.}
         \label{fig:g3e1}
     \end{subfigure}
     ~~~
     \begin{subfigure}{0.4\textwidth}
         \centering
         \includegraphics[width=\textwidth]{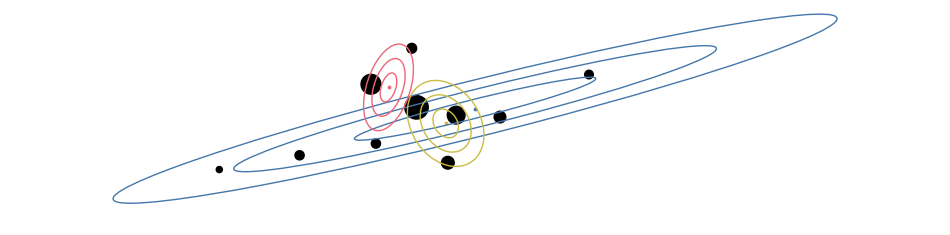}
     \caption{MMD coreset for the sample in Figure~\ref{fig:g3s2}.}
         \label{fig:g3e2}
     \end{subfigure}
     \caption{(a): Three equally weighted Gaussians (lines show 1, 2, 3 standard deviations of each component). (b), (c): Independently sampled datasets from the mixture of three Gaussians. (d), (e): MMD coresets for the three-Gaussian datasets.} 
     \label{fig:three_gaussians}
 \end{figure}

Figure~\ref{fig:three_gaussians} shows that, even if two distributions $X$ and $Y$ are sampled from the same underlying distribution, and their coreset locations are selected from the same collection $U$, the two coresets will not be identical. Here, we see two datasets (Figures~\ref{fig:g3s1} and \ref{fig:g3s2}) generated from the same mixture of three equally weighted Gaussians (Figure~\ref{fig:g3dist}). Below (Figures~\ref{fig:g3e1} and \ref{fig:g3e2}), we have selected a coreset for each dataset (using the algorithm that will be introduced in Section~\ref{sec:alg}), with locations selected from a shared set $U$. While the associated coresets are visually similar, they are not the same.

This is magnified if we look at a high-dimensional dataset. Here, the relative sparsity of data points (and candidate points) in the space means that individual locations in $\mathbb{Q}_X$ might not have close neighbors in $\mathbb{Q}_Y$, even if $X$ and $Y$ are sampled from the same distribution. Further, in high dimensional spaces, it is harder to visually assess the distance between two exemplars. These observations make it hard to compare two coresets, and gain insights about similarities and differences between the associated datasets.
\begin{figure}[h!]
\centering
     \begin{subfigure}[b]{0.4\textwidth}
         \centering
         \includegraphics[width=\textwidth]{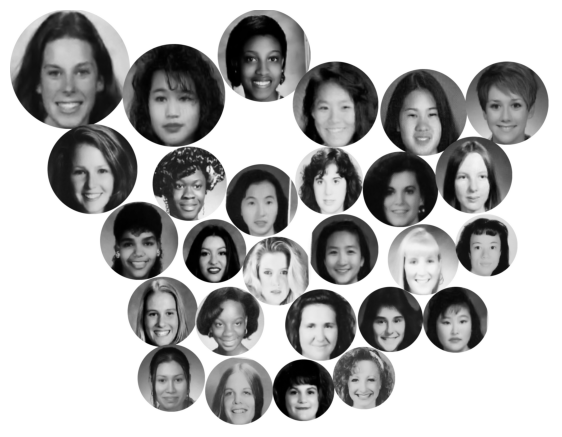}
         \caption{MMD coreset for a set of 250 randomly selected yearbook photos from the 1990s.}
         \label{fig:face_200_90_1a}
     \end{subfigure}
     ~~~
     \begin{subfigure}[b]{0.4\textwidth}
         \centering
         \includegraphics[width=\textwidth]{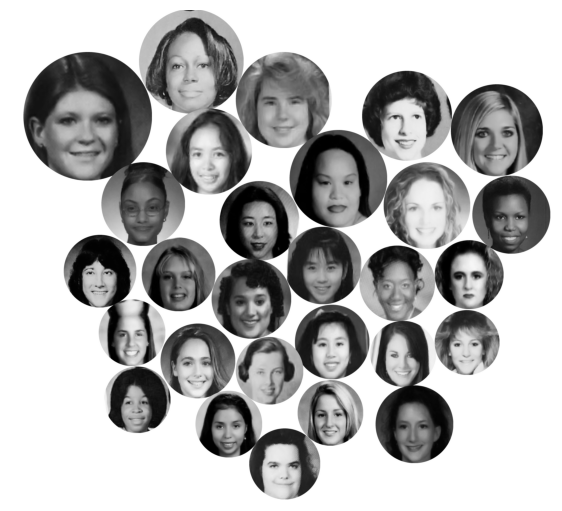}
         \caption{MMD coreset for a second set of 250 randomly selected yearbook photos from the 1990s.}
         \label{fig:face_200_90_2a}
     \end{subfigure}\\
     \caption{Independently learned MMD coresets for two randomly selected dataset of 250 female-identified photographs from US yearbooks in the 1990s. Area of each bubble is proportional to weight of the corresponding exemplar.}\label{fig:basic_face}
\end{figure}
To demonstrate this, we constructed two datasets, each containing 250 randomly selected, female-identified US highschool yearbook photos from the 1990s.
 Figure~\ref{fig:basic_face} shows MMD-coresets obtained for the two datasets.\footnote{See Section~\ref{sec:support_comparison} for full details of dataset and coreset generation.}  While both datasets were selected from the same distribution, there is no overlap in the support of the two coresets. Visually, it is hard to tell that these two coresets are representing samples from the same distribution.

\subsection{Dependent MMD coresets}\label{sec:dependent_mmd_coresets}

The coresets in  Figure~\ref{fig:basic_face} are hard to compare due to their disjoint supports (i.e., the fact that there are no shared exemplars). Comparing the two coresets involves comparing multiple individual photos and assessing their similarities, in addition to incorporating the information encoded in the associated weights. To avoid the lack of interpretability resulting from dissimilar supports, we introduce the notion of a dependent MMD coreset. 

Given a collection of datasets $\{X_t\}_{t\in\mathcal{T}}$, the collection of finite, atomic measures $\{\mathbb{Q}_t\}_{t\in \mathcal{T}}$, is an $\epsilon$-dependent MMD coreset if
\begin{equation}
    \mmdhat(X_t, \mathbb{Q}_t) \leq \epsilon^2 \label{eqn:dep_mmd_cond}
\end{equation}
for all $t\in \mathcal{T}$, and if the $\mathbb{Q}_t$ have common support, i.e.
\begin{equation}
    \mathbb{Q}_t = \sum_{i\in S} w_{t, i} \delta_{u_i},\label{eqn:dep_coreset_form}
\end{equation}
where $\{u_i\}_{i\in S}$ is a subset of some candidate set $U$.

In Equation~\ref{eqn:dep_coreset_form}, the exemplars $u_i$ are shared between all $t\in \mathcal{T}$, but the weights $w_{t, i}$ associated with these exemplars can vary with $t$. Taking the view from Hilbert space, we are restricting the mean embeddings $\mu_{Q_t}$ of the marginal coresets to all lie within a convex hull defined by the exemplars $\{u_i\}_{i\in S}$.

By restricting the support of our coresets in this manner, we obtain data summaries that are easily comparable. Within a single dataset, we can look at the weighted exemplars that make up the coreset and use these to understand the spread of the data, as is the case with an independent MMD coreset. Indeed, since $\mathbb{Q}_t$ still meets the definition of an MMD coreset for $X_t$ (see Equation~\ref{eqn:dep_mmd_cond}), we can use it analogously to an independently generated coreset. However, since the exemplars are shared across datasets, we can directly compare the exemplars for two datasets. We no longer need to intuit similarities between disjoint sets of exemplars and their corresponding weights; instead we can directly compare the weights for each exemplar to determine their relative relevance to each dataset. We will show in Section~\ref{sec:support_comparison} that this facilitates qualitative comparison between the marginal coresets, when compared to independently generated coresets.

We note that the dependent MMD coresets introduced in this paper are directly extensible to other integral probability measures; we could, for example, construct a dependent version of the Wasserstein coresets introduced by \cite{ClaSol2018}.

\subsection{Model understanding and extrapolation}\label{sec:method_understanding}
As we discussed in Section~\ref{sec:bg_understanding},  MMD coresets can be used as tools to understand the performance of an algorithm on ``typical'' data points. Considering how an algorithm performs on such exemplars allows the practitioner to understand failure modes of the algorithm, when applied to the data. In classification tasks where labeling is expensive, or on qualitative tasks such as image modification, looking at an appropriate coreset can provide an estimate of how the algorithm will perform across the dataset.

In a similar manner, dependent coresets can be used to understand generalization behavior of an algorithm. Assume a machine learning algorithm has been trained on a given dataset $X_a$, but we wish to apply it (without modification) to a dataset $X_b$. This is frequently done in practice, since many machine learning algorithms require large training sets and high computational cost; however if the training distribution differs from the deployment distribution, the algorithm may not perform as intended. In general, we would expect the algorithm to perform well on data points in $X_b$ that have many close neighbors in $X_a$, but perform poorly on data points in $X_b$ that are not well represented in $X_a$.

Creating a dependent MMD coreset $\left(\mathbb{Q}_a=\sum_i w_{a, i}\delta_{u_i}, \mathbb{Q}_b =\sum_i w_{b, i}\delta_{u_i}\right)$ for the pair $(X_a, X_b)$ allows us to identify exemplars that are highly representative of $X_a$ or $X_b$ (i.e.\ have high weight in the corresponding weighted coreset). Further, by comparing the weights in the two coreset measures -- e.g, by calculating $f_i = w_{b, i}/w_{a, i}$ -- we can identify exemplars that are much more representative of one dataset than another. Rather than look at all points in the coreset, if we are satisfied with the performance of our model on $X_a$, we can choose to only look at points with high values of $f_i$---points that are representative of the new dataset $X_b$, but not the original dataset $X_a$. Further, if we wish to consider generalization to multiple new datasets, a shared set of exemplars reduces the amount of labeling or evaluation required.

An MMD coreset, dependent or otherwise, will only contain exemplars that are representative of the dataset(s). There are likely to be outliers that are less well represented by the coreset. Such outliers are likely to be underserved by a given algorithm---for example, yielding low accuracy or poor reconstructions. 

As we saw in Section~\ref{sec:bg_mmd_critic}, MMD coresets can be augmented by criticisms---points in $X$ that are poorly approximated by $\mathbb{Q}$. We can equivalently construct criticisms for each dataset represented by a dependent MMD coreset. In the example above, we would select criticisms for the dataset $X_b$ by selecting points in $X_b$ that maximize 
$$\hat{f}_t(x^*) = \frac{1}{n_t}\sum_{i=1}^{n_t} k(x^*, X_{t,i}) - \sum_{j\in S}w_{t,j} k(x^*, u_j)$$ 

In addition to evaluating our algorithm on the marginal dependent coreset for dataset $X_b$, or the subset of the coreset with high values of $f_i$, we can evaluate on the criticisms $C_b$. In conjunction, the dependent MMD coreset and its criticisms allow us to better understand how the algorithm is likely to perform on both typical, and atypical, exemplars of $X_b$.

\section{A greedy algorithm for finding dependent coresets}\label{sec:alg}

Given a collection $\{X_t\}_{t\in\mathcal{T}}$ of datasets, where we assume $X_t:= \{x_{t,1},\dots, x_{t, n_t}\} \sim \mathbb{P}_t$, and a set of $n_U$ candidates $U$, our goal is to find a collection $\{\mathbb{Q}_t\}_{t\in\mathcal{T}}$ with shared support $\{u_i: i\in S\subset [n_U]\}$ such that $\mmdhat(X_t, \mathbb{Q}_t)\leq \epsilon^2$ for all $t\in\mathcal{T}$. We begin by constructing an algorithm for a related task: to minimize $\sum_{t\in\mathcal{T}} MMD(\mathbb{Q}_t, \mathbb{P}_t)$, where

\begin{equation}\mmdhat(X_t, \mathbb{Q}_t) = \frac{1}{n_t^2}\sum_{i=1}^{n_t}\sum_{j=1}^{n_t} k(x_{t,i}, x_{t,j}) + \sum_{i\in S}\sum_{j \in S} w_{t,i} w_{t,j} k(u_i, u_j) - \frac{2}{n_t}\sum_{i=1}^{n_t}\sum_{j\in S} w_{t,j} k(x_i, u_j)\label{eqn:mmd_eqn}\end{equation}
where $\mathbb{Q}_t = \sum_{i\in S} w_{t,i} \delta_{u_i}$. If we ignore terms in Equation~\ref{eqn:mmd_eqn} that do not depend on the $\mathbb{Q}_t$, we obtain the following loss:
\begin{equation}
    \begin{aligned}
    \mathcal{L}(\{\mathbb{Q}_t\}_{t\in\mathcal{T}}) =& \sum_{t\in\mathcal{T}} \ell_t(\mathbb{Q}_t)\\
    \ell_t(\mathbb{Q}_t) =& \frac{1}{2}\sum_{i\in S} \sum_{j\in S} w_{t,i} w_{t,j} k(u_{i}, u_{j}) - \frac{1}{n_t}\sum_{i=1}^{n_t}\sum_{j\in S} k(x_i, u_j)
    \end{aligned}\label{eqn:loss}
\end{equation}

We can use a greedy algorithm to minimize this loss. Let $\mathbb{Q}_t^{(m)} = \sum_{i\in S^{(m)}}w_{t,i}^{(m)} \delta_{u_i^{(m)}}$, where $S^{(m)}$ indexes the first $m$ exemplars to be added. We wish to select the exemplar $u_*$, and set of weights $\left\{w_{t, *}^{(m+1)}, \{w_{t, i}^{(m+1)}\}_{i\in S^{(m)}}\right\}$ for each dataset $X_t$, that minimize the loss. However, searching over all possible combinations of exemplars and weights is prohibitively expensive, as it involves a non-linear optimization to learn the weights associated with each candidate. Instead, we assume that, for each $t\in \mathcal{T}$, there is some $\alpha_t^*>0$ such that $
w_{t, *}^{(m+1)} = \frac{\alpha_t^*}{\alpha_t^*+1}$ and $w_{t, i}^{(m+1)} = \frac{w_{t, i}^{(m)}}{\alpha_t^*+1}$ for all $i\in S^{(m)}$. In other words, we assume that the relative weights in each $\mathbb{Q}_t$ of the previously added exemplars do not change as we add more exemplars.

Fortunately, the value of $\alpha^*_t$ that minimizes $\mathcal{\ell}_t\left( \frac{1}{1+\alpha_t^*}\mathbb{Q}_t^{(m)} + \frac{\alpha_t^*}{\alpha_t^* + 1}\delta_{u_*}\right)$ can be found analytically for each candidate $u_*$ by differentiating the loss in step 1, yielding

\begin{equation}\alpha_t^* = \frac{ \displaystyle \sum_{i, j\in S^{(m)}} w_{t, i}w_{t, j}k(u_i, u_j) - \sum_{i\in S^{(m)}} w_{t,i} k(u_i, u_*) - \frac{1}{n_t}\sum_{i=1}^{n_t}\left(k(x_i, u_*) - \sum_{j\in S^{(m)}} w_j k(x_{t,i}, u_j)\right)}{\displaystyle k(u_*, u_*) - \sum_{i\in S^{(m)}} w_{t,i} k(u_i, u_*) - \frac{1}{n_t}\sum_{i=1}^{n_t}\left(k(x_i, u_*) - \sum_{j\in S^{(m)}} w_j k(x_{t,i}, u_j) \right)}.\label{eqn:opt_alpha}\end{equation}

We can therefore set 
\begin{equation}i^*, \{\alpha_t^*\} \leftarrow \argmin_{\stackrel{i\in [n_U]\setminus S^{(m)},}{\alpha_t \in \mathbb{R}_+}} \sum_{t\in\mathcal{T}}\ell_t\left( \frac{1}{1+\alpha_t}\mathbb{Q}_t^{(m)} + \frac{\alpha_t}{\alpha_t + 1}\delta_{u_i}\right)\label{eqn:optimize}\end{equation}
and let $S^{(m+1)}=S^{(m)}\cup i^*$, $w^{(m+1)}_{t, i^*}=\frac{\alpha_{t}^*}{\alpha_t^*+1}$ for all $t\in \mathcal{T}$, and $w^{(m+1)}_{t, i} = \frac{w_{t,i}^{(m)}}{\alpha_t^*+1}$ for all $t\in\mathcal{T}$ and $i\in S^{(m)}$.

As written, the procedure will greedily minimize the sum of the per-dataset losses. However, the definition of an MMD coreset involves satisficing, not minimizing: we want $\mmdhat(X_t, \mathbb{Q}_t)\leq \epsilon^2$ for all $t\in \mathcal{T}$.  To achieve this, we modify the sum in Equation~\ref{eqn:optimize} so that it only includes terms for which  $\mmdhat(X_t, \mathbb{Q}_t^{(m)})>\epsilon^2$. The resulting procedure is summarized in Algorithm~\ref{alg:dmmd}.

\begin{algorithm} % enter the algorithm environment
\caption{{\sc{dmmd}}: Selecting dependent MMD coresets} % give the algorithm a caption
\label{alg:dmmd} % and a label for \ref{} commands later in the document
\begin{algorithmic} % enter the algorithmic environment
    \REQUIRE Datasets $\{X_t\}_{t\in \mathcal{T}}$; candidate set $U$; kernel $k(\cdot, \cdot)$; threshold $\epsilon^2>0$
    \STATE $S^{(0)} \leftarrow \emptyset$; $w_t^{(0)} \leftarrow [\,]$ for all $t\in\mathcal{T}$; $D\leftarrow \mathcal{T}$; $m\leftarrow 0$
    \WHILE{$D\neq \emptyset$}
        \FORALL{$i \in [n_U]\setminus S^{(m)}$}
            \FORALL{$t\in\mathcal{T}$}
                \STATE{Calculate $\alpha^*_{t, i}$ using Equation~\ref{eqn:opt_alpha}}
            \ENDFOR 
            \STATE{$L_{i} \leftarrow 0$}
            \FORALL{$t\in D$}
                \STATE{$L_i \leftarrow L_i + \ell_t\left( \frac{1}{1+\alpha_t^*}\mathbb{Q}_t^{(m)} + \frac{\alpha_t^*}{\alpha_t^* + 1}\delta_{u_*}\right)$}
            \ENDFOR
        \ENDFOR
    \STATE{$i^* = \argmin_{i\in [n_U]\setminus S^{(m)}} L_i$}
    \STATE{$S^{(m+1)}\leftarrow S^{(m)} \cup \{i^*$\}}
    \STATE{$w_{i^*}^{(m+1)}\leftarrow \frac{\alpha_{i^*}}{\alpha_{i^*}+1}$}
    \FORALL{$t\in\mathcal{T}$}
        \FORALL{$i\in S^{(m)}$}
            \STATE{$w_i^{(m+1)} \leftarrow \frac{w_i^{(m)}}{\alpha_{i^*}+1}$}
        \ENDFOR
        \STATE{$\mathbb{Q}_t^{(m+1)}\leftarrow \sum_{i\in S^{(m+1)}}w_{t, i}^{(m+1)}\delta_{u_i}$}
    \ENDFOR
    \STATE{$D = \{X_t: \mmdhat(X_t, \mathbb{Q}_t^{(m+1)}) > \epsilon^2\}$}
    \STATE{$m\leftarrow m+1$}
    \ENDWHILE
\end{algorithmic}
\end{algorithm}

\section{Experimental evaluation}\label{sec:experiments}

We begin in Section~\ref{sec:results_alg} by evaluating performance of Algorithm~\ref{alg:dmmd}. Next, we show, using image datasets as an example, how dependent MMD coresets can be used to understand differences between related datasets (Section~\ref{sec:results_interpretable}) and gain insights about model generalization (Section~\ref{sec:results_model_understanding}).
\subsection{Evaluation of dependent MMD coreset algorithm}\label{sec:results_alg}
We begin by evaluating the performance of our algorithm (which we denote {\sc{dmmd}}) for finding dependent MMD coresets, compared with two alternative methods.

\paragraph{{\sc{protodash}}} The {\sc{protodash}} algorithm \cite{GurDhuCecAgg2019} for weighted MMD coresets greedily selects exemplars that minimize the gradient of the loss in Equation~\ref{eqn:loss} (for  a single dataset). Having selected an exemplar to add to the coreset, {\sc{protodash}} then uses an optimization procedure to find the weights that minimize $\mmdhat(X, \mathbb{Q}^{(m+1)})$. We modify this algorithm for the dependent MMD setting by summing the gradients across all datasets for which the $\epsilon^2$ threshold is not yet satisficed, leading to the dependent {\sc{protodash}} algorithm shown in Algorithm~\ref{alg:protodash}.

\paragraph{\sc dmmd-opt}Unlike the dependent version of \texttt{protodash} in Algorithm~\ref{alg:protodash}, our algorithm assigns weights \textit{before} selection, which should encourage adding points that would help some of the marginal coresets, but not others. However, there is no post-exemplar-addition optimization of the weights. Inspired by the post-addition optimization in {\sc{protodash}}, we also compare our algorithm with a variant of Algorithm~\ref{alg:dmmd} that optimizes the weights after each step---allowing the relative weights of the exemplars to change between each iteration. We will refer to this variant of {\sc{dmmd}} with post-exemplar-addition optimization as {\sc{dmmd-opt}}.

\begin{algorithm} % enter the algorithm environment
\caption{A dependent \texttt{protodash} algorithm} % give the algorithm a caption
\label{alg:protodash} % and a label for \ref{} commands later in the document
\begin{algorithmic} % enter the algorithmic environment
    \REQUIRE Datasets $\{X_t\}_{t\in \mathcal{T}}$, candidate set $U$, kernel $k(\cdot, \cdot)$, threshold $\epsilon^2>0$
    \STATE $S^{(0)} \leftarrow \emptyset$, $w_t^{(0)} \leftarrow []$ for all $t\in\mathcal{T}$, $D\leftarrow \mathcal{T}$, $m\leftarrow 0$
    \FORALL{$i\in [n_U]$}
        \STATE{$g_{i} =  \sum_{t\in\mathcal{T}}\frac{1}{n_t}\sum_{j=1}^{n_t} k(x_j, u_i)$}
    \ENDFOR
    \WHILE{$D\neq \emptyset$}
        \STATE{$i^* = \argmin_{i\in [n_U]\setminus S^{(m)}} g_i$}
        \STATE{$S^{(m+1)}\leftarrow S^{(m)}\cup \{i^*\}$}
        \FORALL{$t\in\mathcal{T}$}
            \STATE{$\{w_{t, i}^{(m+1)}\}_{i\in S^{(m+1)}} \leftarrow \argmax_{\{w_{t,i}\}_{i\in S^{(m+1)}}} \ell_t(\sum_{i\in S^{(m+1)}}w_{t,i}\delta_{u_i})$}
        \ENDFOR
        \STATE{$\mathbb{Q}_t^{(m+1)}\leftarrow \sum_{i\in S^{(m+1)}}w_{t, i}^{(m+1)}\delta_{u_i}$}
        \FORALL{$i\in [n_U]\setminus S^{(m+1)}$}
            \STATE{$g_i = \sum_{t\in D} \nabla \ell_t(\mathbb{Q}^{(m+1)}$}
        \ENDFOR
        \STATE{$D = \{X_t: \mmdhat(X_t, \mathbb{Q}_t^{(m+1)}) > \epsilon^2\}$}
        \STATE{$m\leftarrow m+1$}
    \ENDWHILE
\end{algorithmic}
\end{algorithm}

\begin{figure}
    \centering
    \includegraphics[width=.4\textwidth]{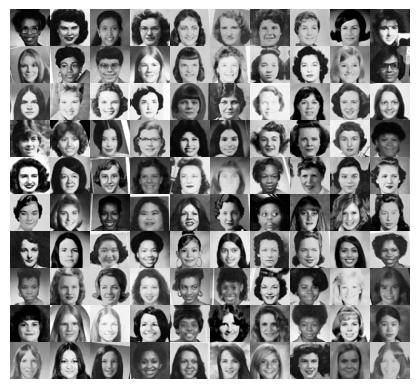}
    \caption{A random subset of 100 images taken from the yearbook dataset.}
    \label{fig:random_faces}
\end{figure}

We evaluate all three methods using a dataset of photographs of 15367 female-identified students, taken from yearbooks between 1905 and 2013 \cite{GinRakSachYinEfr2015}. We show a random subset of these images in Figure~\ref{fig:random_faces}. We generated 512-dimensional embeddings of the photos using the torchvision pre-trained implementation of ResNet \cite{marcel2010torchvision,resnet}. We then partitioned the collection into 12 datasets, each containing photos from a single decade.

 In order to capture lengthscales appropriate for the variation in each decade, we use an additive kernel, setting
$$K = \frac{K_{all} + \sum_{t\in\mathcal{T}} K_t}{|\mathcal{T}| + 1},$$
where $K_{all}$ is a squared exponential kernel with bandwidth given by the overall median pairwise distances; $\mathcal{T}$ is the set of decades that index the datasets; $K_t$ is a squared exponential kernel with bandwidth given by the median pairwise distance between images in dataset $t$. 

We begin by considering how good a dependent MMD coreset each algorithm is able to construct, for a given number of exemplars $m=|S|$. To do so, we ran all algorithms without specifying a threshold $\epsilon^2$, recording $\mmdhat(X_t, \mathbb{Q}_t^{(m)})$ for each value of $m$. All algorithms were run for one hour on a 2019 Macbook Pro (2.6 GHz 6-Core Intel Core i7, 32 GB 2667 MHz DDR4), excluding time taken to generate and store the kernel entries, which only occurs one time. As much code as possible was re-used between the three algorithms. Where required, optimization of weights was carried out using a BFGS optimizer. Code is available at \url{https://github.com/sinead/dmmd}.

\begin{figure}
    \centering
    \begin{subfigure}{0.7\textwidth}
     \includegraphics[width=\textwidth]{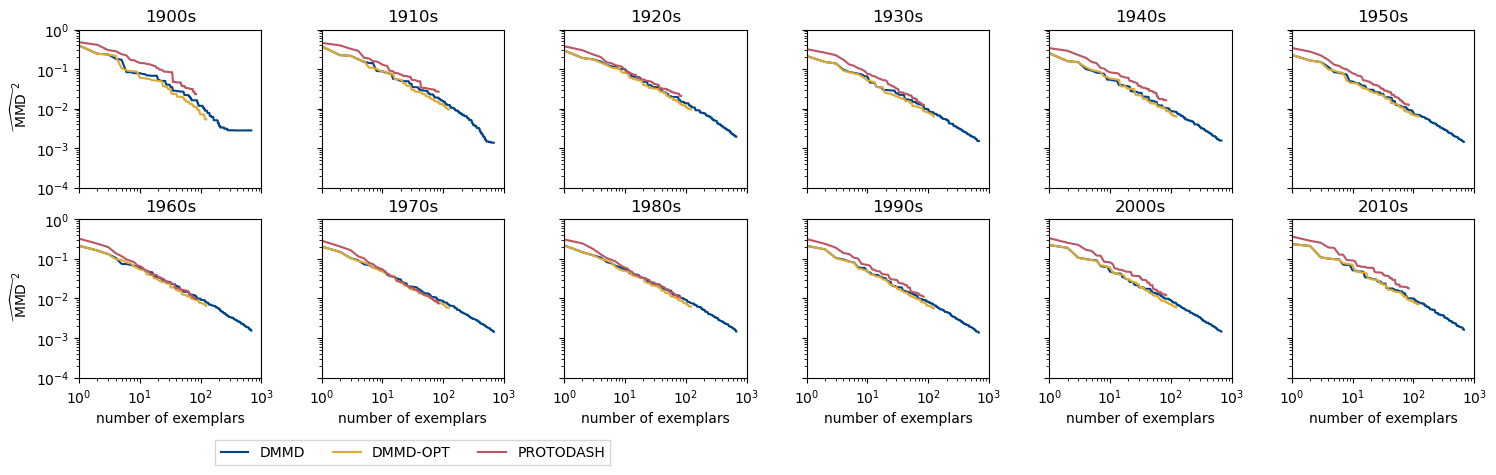}
    \caption{$\mmdhat(X_t, \mathbb{Q}_t^{(m)}$ for increasing numbers of exemplars $m$. Each plot corresponds to a dataset containing yearbook photos from a single decade.}
    \label{fig:indiv_mmd_vs_m}
    \end{subfigure}\\
    \vspace{.1in}
   \begin{subfigure}{0.7\textwidth}
   \centering
   \includegraphics[width=.4\textwidth]{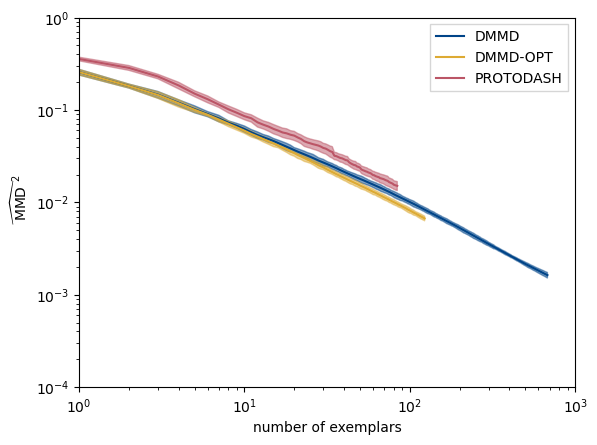}
    \caption{Mean $\pm$ one standard error of $\mmdhat(X_t, \mathbb{Q}_t^{(m)}$ for increasing numbers of exemplars $m$. }
    \label{fig:av_mmd_vs_m}
   \end{subfigure}
   \caption{Evaluating how coreset quality varies with number of exemplars, for dependent MMD coresets generated using three algorithms, on 12 yearbook datasets.}
\end{figure}

Figure~\ref{fig:indiv_mmd_vs_m} shows the per-dataset estimates $\mmdhat(X_t, \mathbb{Q}_t^{(m)})$, and Figure~\ref{fig:av_mmd_vs_m} shows the average performance across all 12 datasets. We see that the three algorithms perform comparably in terms of coreset quality. {{\sc{dmmd-opt}}} seems to perform slightly better than {\sc{dmmd}}, as might be expected due to the additional optimization step. {\sc{protodash}}, by comparison, seems to perform slightly worse, which we hypothesise is because it has no mechanism by which weights can be incorporated at selection time. However, in both cases, the difference is slight.

\begin{figure}
    \centering
    \includegraphics[width=.3\textwidth]{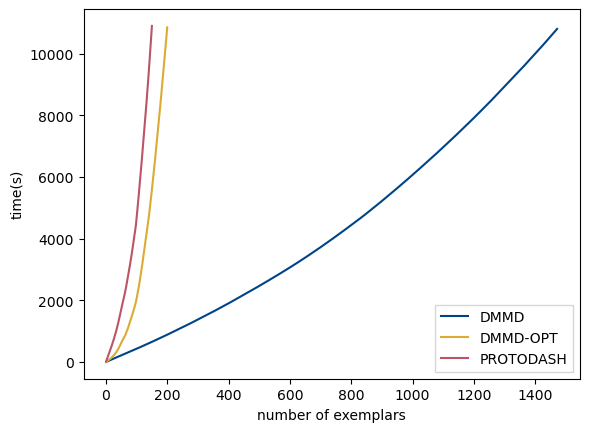}
    \caption{Time (in seconds) taken to construct MMD dependent coresets of a given size, for three algorithms, on the 12 yearbook datasets. Algorithms ran for a maximum of one hour.}
    \label{fig:m_vs_time}
\end{figure}

{\sc{dmmd}} is however \textit{much} faster at generating coresets, since it does not optimize the full set of weights at each iteration. This can be seen in Figure~\ref{fig:m_vs_time}, which shows the time taken to generate coresets of a given size. The cost of the optimization-based algorithms grows rapidly with coreset size $(m)$; the rate of growth of the {\sc{dmmd}} coresets is much smaller.

\begin{figure}
    \centering
    \includegraphics[width=.3\textwidth]{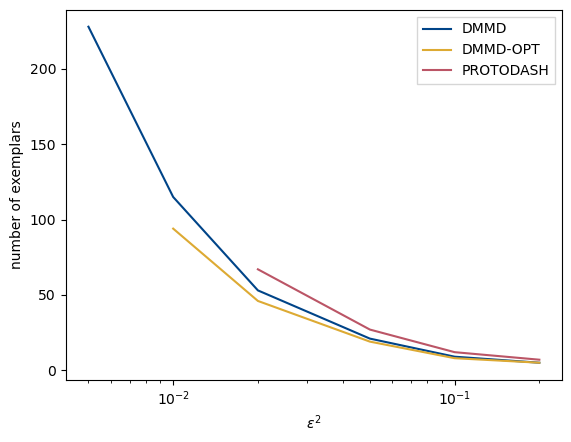}
    \caption{Coreset size required to obtain an $\epsilon$-MMD dependent coreset on the 12 yearbook datasets, for three algorithms. Algorithms ran for a maximum of one hour; {\sc{ protodash}} failed to complete coresets for $\epsilon^2=0.01$ and $\epsilon^2=0.005$. {\sc{dmmd-opt}} failed to complete a coreset for $\epsilon^2=0.005$.}
    \label{fig:m_vs_thresh}
\end{figure}

In practice, rather than endlessly minimizing $\sum_{t\in\mathcal{T}}\mmdhat(X_t, \mathbb{Q}_t)$, we will aim to find $\mathbb{Q}_t$ such that $\mmdhat(X_t, \mathbb{Q}_t)<\epsilon^2$ for all $t\in\mathcal{T}$. In Figure~\ref{fig:m_vs_thresh}, we show the coreset sizes required to obtain an $\epsilon$-MMD dependent coreset on the twelve decade-specific yearbook datasets, for each algorithm. Again, a maximum runtime of one hour was specified. When all three algorithms were able to finish, the coreset sizes are comparable (with {\sc{dmmd-opt}} finding slightly smaller coresets than {\sc{dmmd}}, and {\sc{protodash}} finding slightly larger coresets). However the optimization-based methods are hampered by their slow runtime.

Based on these analyses, it appears there is some advantage to additional optimization of the weights. However, in most cases, we do not feel the additional computational cost merits the improved performance.

\subsection{Interpretable data summarizations}\label{sec:results_interpretable}
In high-dimensional, highly structured datasets such as collections of images, traditional summary statistics such as the mean of a dataset are particularly uninterpretable, as they convey little of the shape of the underlying distribution. 
A better approach is to show the viewer a collection of images that are representative of the dataset. MMD coresets allow us to obtain such a representative set, making them a better choice than displaying a random subset.

As we discussed in Section~\ref{sec:limitations}, if we wish to summarize a collection of related datasets, independently generated MMD coresets can help us understand each dataset individually, but it may prove challenging to compare datasets. This challenge becomes greater in high dimensional settings such as image data, where we cannot easily intuit a distance between exemplars. To showcase this phenomenom, and demonstrate how dependent MMD coresets can help, we return to the 15367 yearbook photos introduced in Section~\ref{sec:results_alg}. For all experiments in this section, we use the additive kernel described in Section~\ref{sec:results_alg}.

\begin{figure}[t]
\centering
     \begin{subfigure}[b]{0.4\textwidth}
         \centering
         \includegraphics[width=\textwidth]{figures/mmd_90s_1.png}
         \caption{0.1-MMD coreset for a set of 250 randomly selected yearbook photos from the 1990s.}
         \label{fig:face_200_90_1}
     \end{subfigure}
     ~~~
     \begin{subfigure}[b]{0.4\textwidth}
         \centering
         \includegraphics[width=\textwidth]{figures/mmd_90s_2.png}
         \caption{0.1-MMD coreset for a second set of 250 randomly selected yearbook photos from the 1990s.}
         \label{fig:face_200_90_2}
     \end{subfigure}\\
     \vspace{0.05in}
     
     \begin{subfigure}[b]{0.4\textwidth}
         \centering
         \includegraphics[width=\textwidth]{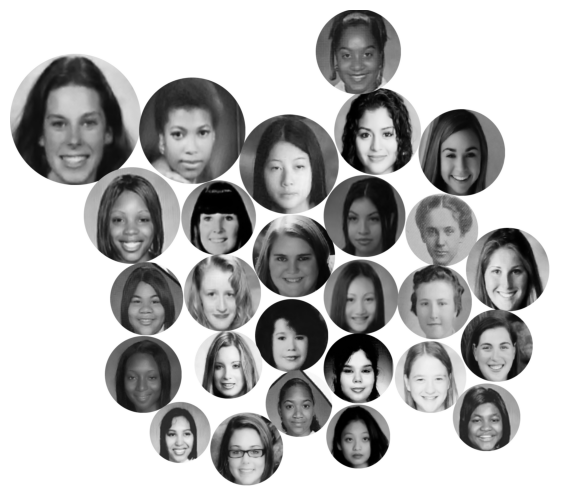}
         \caption{0.1-MMD coreset for a set of 250 randomly selected yearbook photos from the 2000s.}
         \label{fig:face_200_00_1}
     \end{subfigure}
     ~~~
     \begin{subfigure}[b]{0.4\textwidth}
         \centering
         \includegraphics[width=\textwidth]{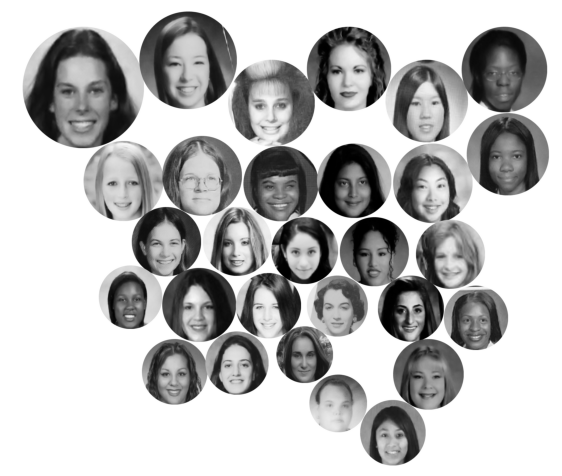}
         \caption{0.1-MMD coreset for a second set of 250 randomly selected yearbook photos from the 2000s.}
         \label{fig:face_200_00_2}
     \end{subfigure}
     \caption{Independently learned, weighted 0.1-MMD coresets based on 250 random samples from a given decade. Area of each bubble is proportional to the weight of the corresponding exemplar in the coreset.}\label{fig:face_200_exemplars}
\end{figure}
\subsubsection{A shared support allows for easier comparison}\label{sec:support_comparison}
In Section~\ref{sec:dependent_mmd_coresets}, we argued that the shared support provided by dependent MMD coresets facilitates comparison of datasets, since we only need to consider differences in weights. To demonstrate this, we constructed four datasets, each a subset of the entire yearbook dataset containing 250 photos. The first two datasets contained only faces from the 1990s; the second two, only faces from the 2000s. The datasets were generated by sampling without replacement from the associated decades, to ensure no photo appeared more than once across the four datasets.

We begin by independently generating MMD coresets for the four datasets, using Algorithm~\ref{alg:dmmd} independently on each dataset, with a threshold of $\epsilon^2 = 0.01$. The set of candidate images, $U$, was the entire dataset of 15367 images. The resulting coresets are shown in Figure~\ref{fig:face_200_exemplars}; the areas of the bubbles correspond to the weights associated with each exemplar.\footnote{The top row of Figure~\ref{fig:face_200_exemplars} duplicates Figure~\ref{fig:basic_face}.}

We can see that, considered individually, each coreset appears to be doing a good job of capturing the variation in students for each dataset. However, if we compare the four coresets, it is not easy to tell that Figures~\ref{fig:face_200_90_1} and \ref{fig:face_200_90_2} represent the same underlying distribution, and Figures~\ref{fig:face_200_00_1} and \ref{fig:face_200_00_2} represent a second underlying distribution---or to interpret the difference between the two distributions. We see that the highest weighted exemplar for the two 2000s datasets is the same(top left of Figures~\ref{fig:face_200_00_1} and \ref{fig:face_200_00_2}), but only one other image is shared between the two coresets. Meanwhile, the first coreset for the 1990s shares the same highest-weighted image with the two 2000s datasets---but this coreset does not appear in the first 1990s coreset, and the two 1990s coresets have no overlap.  Overall, it is hard to compare between the marginal coresets.

\begin{figure}[t]
\centering
     \begin{subfigure}[b]{0.4\textwidth}
         \centering
         \includegraphics[width=\textwidth]{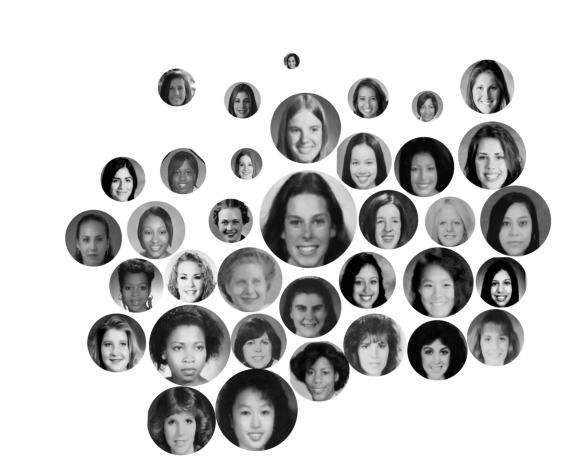}
         \caption{Marginal dependent 0.1-MMD coreset for a set of 250 randomly selected yearbook photos from the 1990s.}
         \label{fig:dep_face_200_90_1}
     \end{subfigure}
     ~~~
     \begin{subfigure}[b]{0.4\textwidth}
         \centering
         \includegraphics[width=\textwidth]{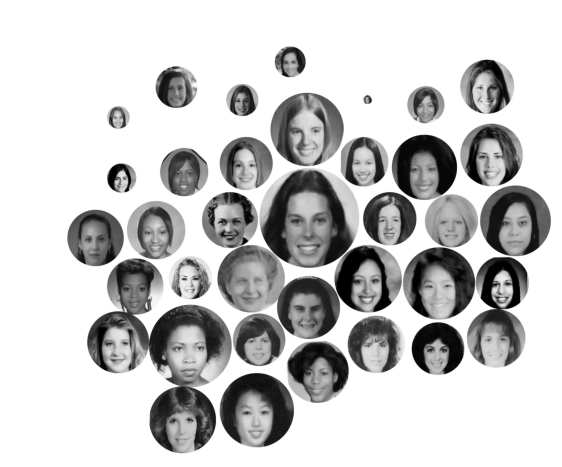}
         \caption{Marginal dependent 0.1-MMD coreset for a second set of 250 randomly selected yearbook photos from the 1990s.}
         \label{fig:dep_face_200_90_2}
     \end{subfigure}\\
     \vspace{0.05in}
     
     \begin{subfigure}[b]{0.4\textwidth}
         \centering
         \includegraphics[width=\textwidth]{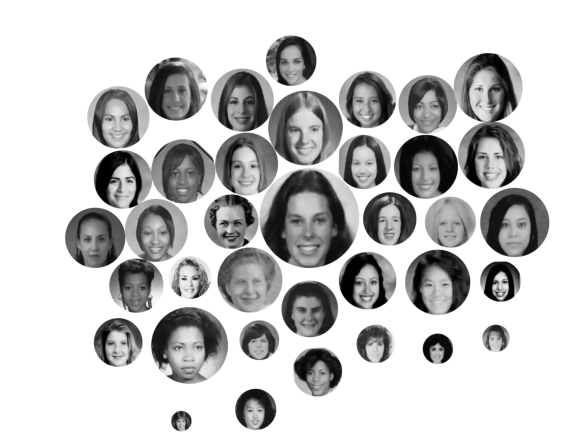}
         \caption{Marginal dependent 0.1-MMD coreset for a set of 250 randomly selected yearbook photos from the 2000s.}
         \label{fig:dep_face_200_00_1}
     \end{subfigure}
     ~~~
     \begin{subfigure}[b]{0.4\textwidth}
         \centering
         \includegraphics[width=\textwidth]{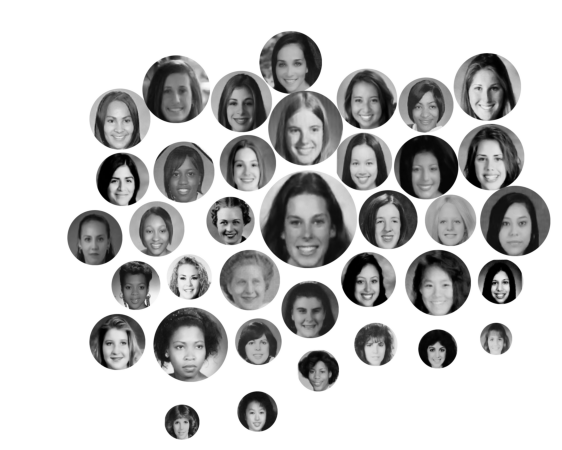}
         \caption{Marginal dependent 0.1-MMD coreset for a second set of 250 randomly selected yearbook photos from the 2000s.}
         \label{fig:dep_face_200_00_2}
     \end{subfigure}
     \caption{Dependent 0.1-MMD coreset for a collection of 4 datasets, each including 250 random samples from a given decade. Area of each bubble is proportional to the weight of the corresponding exemplar in the marginal coreset. Positioning is constant across all four examples.}\label{fig:dep_face_200_exemplars}
\end{figure}

By contrast, the shared support offered by dependent coresets means we can directly compare the distributions using their coresets. In Figure~\ref{fig:dep_face_200_exemplars}, we show a dependent MMD coreset ($\epsilon^2 = 0.01$) for the same collection of datasets. The shared support allows us to see that, while the two decades are fairly similar, there is clearly a stronger similarity between the pairs of datasets from the same year (i.e., similarly sized photos), than between pairs from different years. We can also identify images that exemplify the difference between the two decades, by looking at the difference in weights. We see that many of the faces towards the top of the bubble plot have high weights in the 2000s, but low weights in the 1990s. Examining these exemplars suggests that straight hair became more prevalent in the 2000s. Conversely, many of the faces towards the bottom of the bubble plot have high weights in the 1990s, but low weights in the 2000s. These photos tend to have wavy/fluffy hair and bangs. In conjunction, these plots suggest a tendency in the 2000s away from bangs and towards straight hair, something the authors remember from their formative years. However, there is still a significant overlap between the two decades: many of the exemplars have similar weights in the 1990s and the 2000s. 

\begin{figure}[t]
     \centering
     \includegraphics[width=\textwidth]{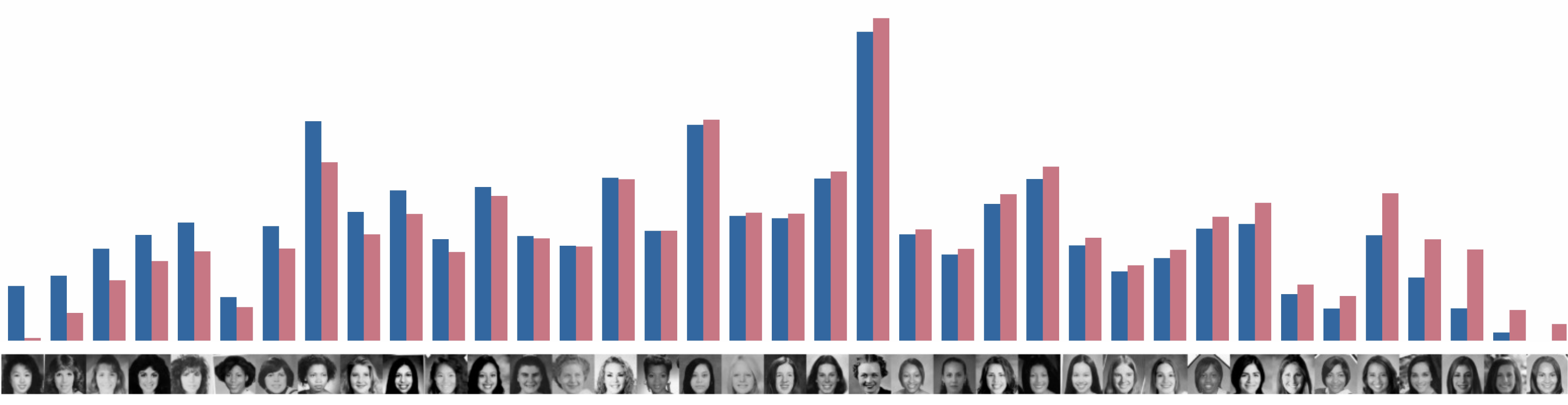}
     \caption{Summary of a dependent 0.1-MMD coreset for four datasets of yearbook faces from the 1990s and 2000s. Exemplars are shown along the $x$ axis. The average weight for each exemplar in the coresets associated with the 1990s is shown in blue {\color{bluetol}\rule{\widthof{block}}{\heightof{block}}}; the average weight for the 2000s is shown in red {\color{redtol}\rule{\widthof{block}}{\heightof{block}}.\label{fig:yearbook_hist}}}
\end{figure}

We can also see this in Figure~\ref{fig:yearbook_hist}, a bar chart shows the average weights associated with each exemplar in each decades (i.e., the blue bar above a given image is the average weight for that exemplar across the two 1990s datasets, and the red bar is the average weight across the two 2000s datasets). We see that most of the exemplars have similar weights in both scenarios, but that we have a number of straight-haired exemplars disproportionally representing the 2000s, and a number of exemplars with bangs and/or wavy hair disproportionately representing the 1990s.

\subsubsection{Dependent coresets allow us to visualize data drift in collections of images}
Next, we show how dependent MMD coresets can be used to understand and visualize variation between collections of multiple datasets. As in Section~\ref{sec:results_alg}, we partition the 15367 yearbook images based on their decade, with the goal of understanding how the distribution over yearbook photos changes over time. Table~\ref{tab:decade_counts} shows the number of photos in each resulting dataset.

\begin{table}[h]
    \centering
    \begin{tabular}{c c c c c c c c c c c c}
1900s & 1910s & 1920s & 1930s & 1940s & 1950s & 1960s & 1970s & 1980s & 1990s & 2000s & 2010s\\
\hline
35 & 98 & 308 & 1682 & 2650 & 2093 & 2319 & 2806 & 2826 & 2621 & 2208 & 602
    \end{tabular}
    \caption{Number of yearbook photos for each decade.}
    \label{tab:decade_counts}
\end{table}

\begin{figure}[t]
     \centering
     \includegraphics[width=.9\textwidth]{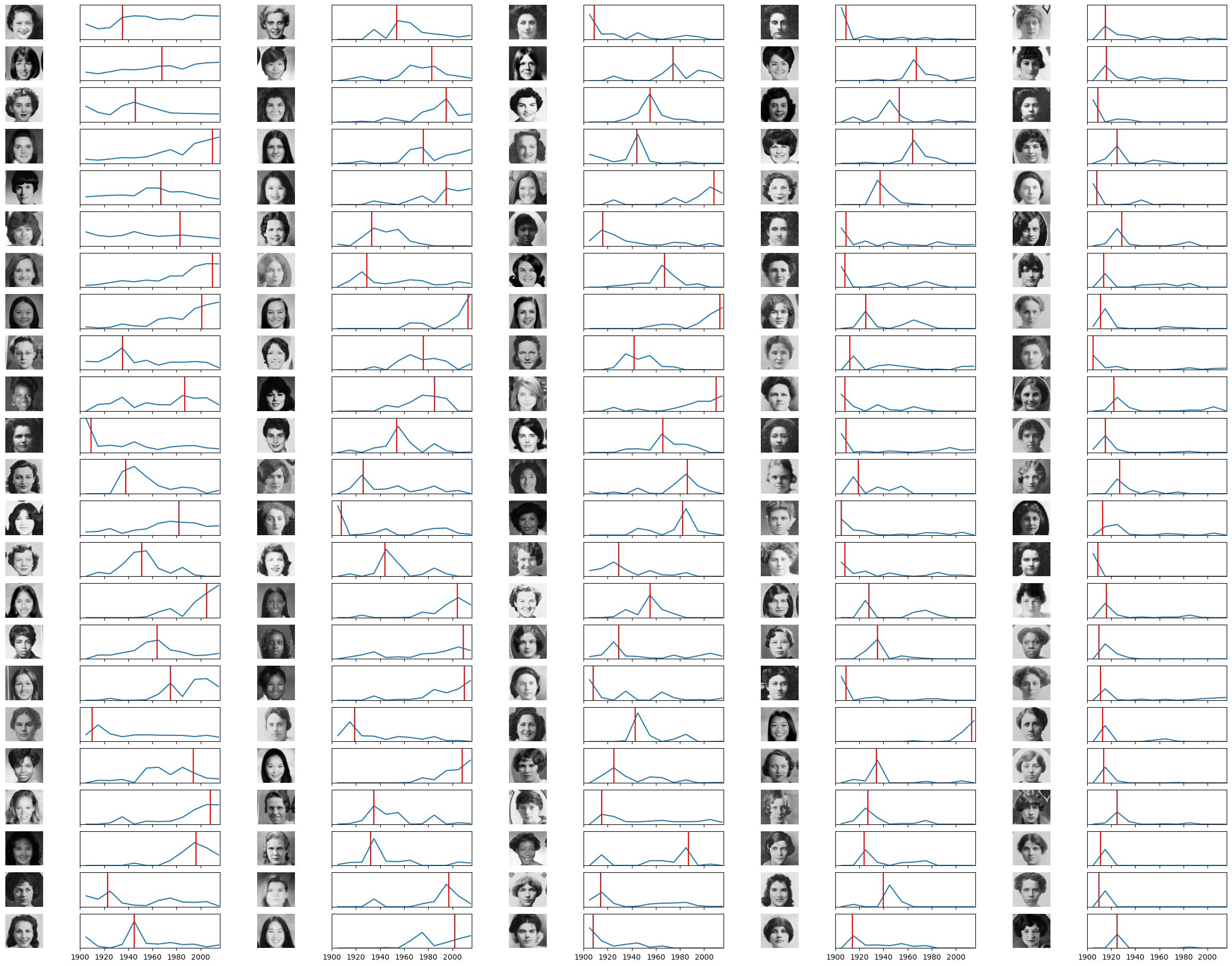}
     \caption{Visualization of an 0.1-MMD dependent coreset for 12 datasets, each containing yearbook photos from a given decades. Photos show the exemplars $\{u_i: i\in S\}$, ordered by their average weight across the 12 marginal coresets. To the left of each photo is a plot of the corresponding weight over time; a red vertical line marks the year the photo was taken.\label{fig:yearbook_lines}}
\end{figure}

\begin{figure}[h]
    \centering
    \includegraphics[width=.4\textwidth]{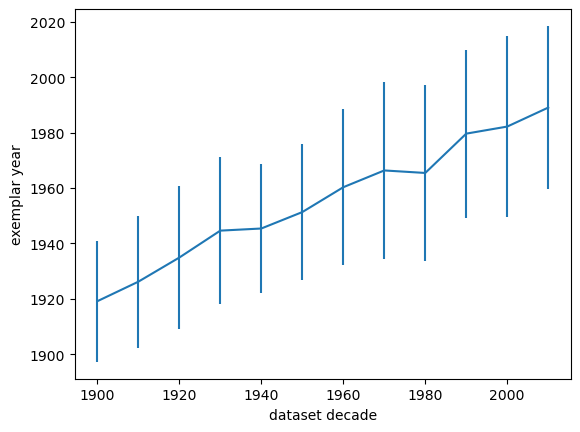}
    \caption{Distribution over the year associated with the marginal coresets for each decade. Plot shows weighted mean $\pm$ weighted standard deviation.}
    \label{fig:weighted_mean_1}
\end{figure}
Figure~\ref{fig:yearbook_lines} shows the resulting exemplars with a threshold of $\epsilon^2=0.01$. The corresponding plots show how the weights vary with time. The exemplars are ordered based on their average weight across the 12 datasets. In each case, a red, vertical line indicates the year of the yearbook from which the exemplar was taken. We are able to see how styles change over time, moving away from the formal styles of the early 20th century, through waved hairstyles popular in the midcentury, towards longer, straighter hairstyles in later decades. In general, the relevance of an exemplar peaks around the time it was taken (although, this information is not used to select exemplars). However, some styles remain relevant over longer time periods (see many exemplars in the first column). Most of the early exemplars are highly peaked on the 00s or 10s; this is not surprising since these pre-WW1 photos tend to have very distinctive photography characteristics and hair styles.

Figure~\ref{fig:yearbook_lines} appears to show that the marginal coresets have high weights on exemplars from the corresponding decade. To look at this in more detail, we consider the distributions over the dates of the exemplars associated with each decade. Figure~\ref{fig:weighted_mean_1} shows the weighted mean and standard deviation of the years associated with the exemplars, with weights given by the coreset weights.  We see that the mean weighted year of the exemplars increases with the decade. However, we notice that it is pulled towards the 1940s and 1950s in each case: this is because we must represent all datasets using a weighted combination of points taken from the convex hull of all datapoints.

\subsection{Dependent coresets allow us to understand model generalization}\label{sec:results_model_understanding}

\begin{figure}[h]
    \centering
    \begin{subfigure}{0.3\textwidth}
    \includegraphics[width=.9\textwidth]{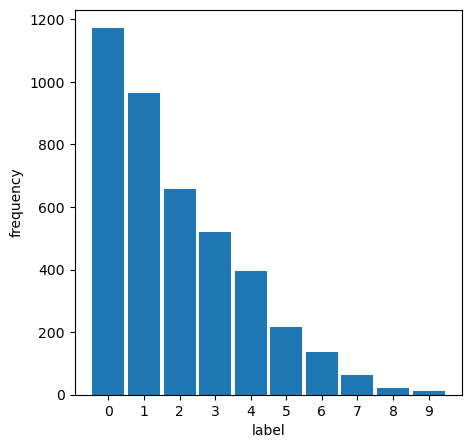}
        \caption{Label frequencies for dataset $X_a$}
    \end{subfigure}
    ~~~
    \begin{subfigure}{0.3\textwidth}
    \includegraphics[width=.9\textwidth]{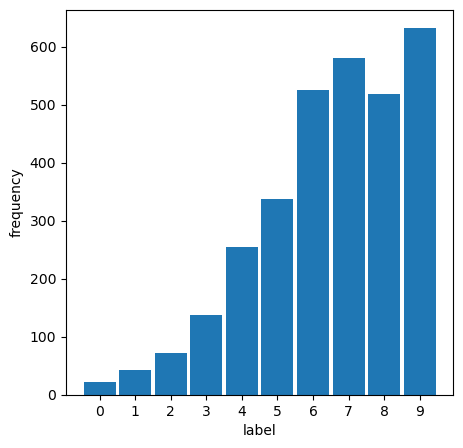}
        \caption{Label frequencies for dataset $X_b$}
    \end{subfigure}
    \caption{Frequency with which each digit occurs in two datasets of handwritten digits}
    \label{fig:number_hist}
\end{figure}

To see how dependent coresets can be used to understand how a model trained on one dataset will generalize to others, we create two datasets of image digits. We started with the USPS handwritten digits dataset \cite{uspsdataset}, which comprises a train set of 2791 handwritten digits, and a test set of 2001 handwritten digits. We split the train set into two datasets, $X_a$ and $X_b$, where $X_a$ is skewed towards the earlier digits and $X_b$ towards the later digits. Figure~\ref{fig:number_hist} shows the resulting label counts for each dataset.

Our goal is to use dependent MMD coresets to qualitatively assess how well algorithms trained on $X_a$ will generalize to $X_b$. To ensure the exemplars in our coreset have not been seen in training, we let our set of candidate points $U$ be the union of $X_b$ and the USPS test set. As with the yearbook data, we use an additive squared exponential kernel, with bandwidths of the composite kernels being the median within-class pairwise distances, and the overall median pairwise distance. Distances were calculated using the raw pixel values.

\begin{figure}[ht]
    \centering
    \includegraphics[width=\textwidth]{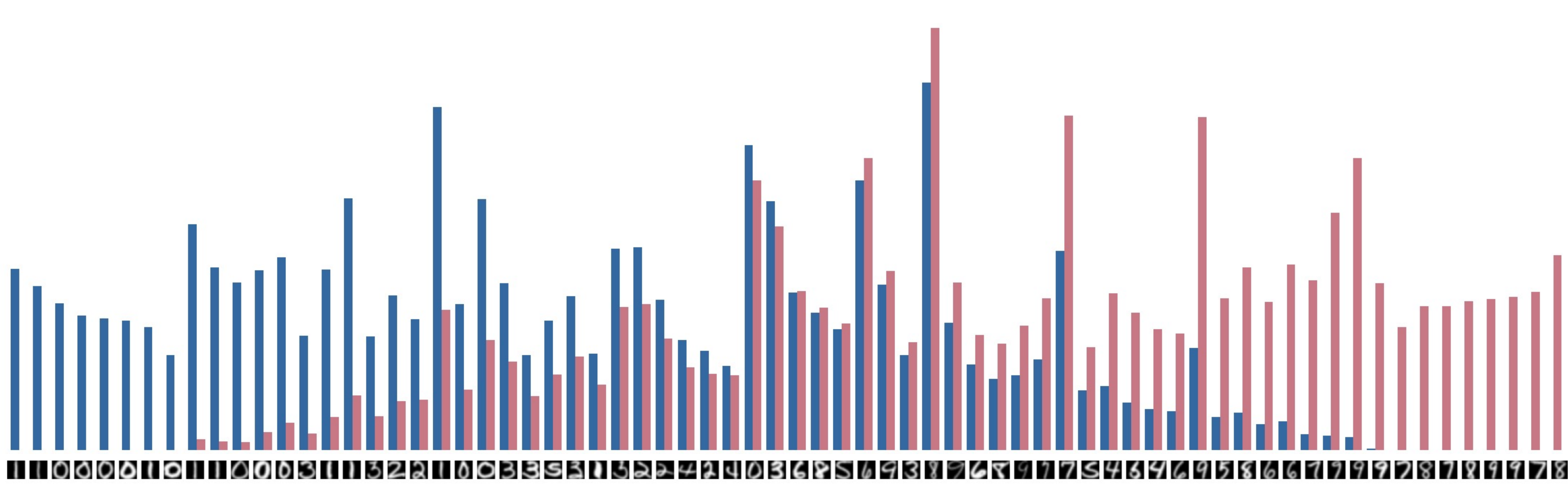}
    \caption{Dependent MMD coreset for two datasets of handwritten digits. Bars show the weight in each marginal coreset, with $X_a$ shown in blue {\color{bluetol}\rule{\widthof{block}}{\heightof{block}}} and $X_b$ shown in red {\color{redtol}\rule{\widthof{block}}{\heightof{block}}}; images along axis show corresponding exemplars.
    \label{fig:usps_hist_1}}
\end{figure}

\begin{figure}[h!]
    \centering
    \includegraphics[width=.3\textwidth]{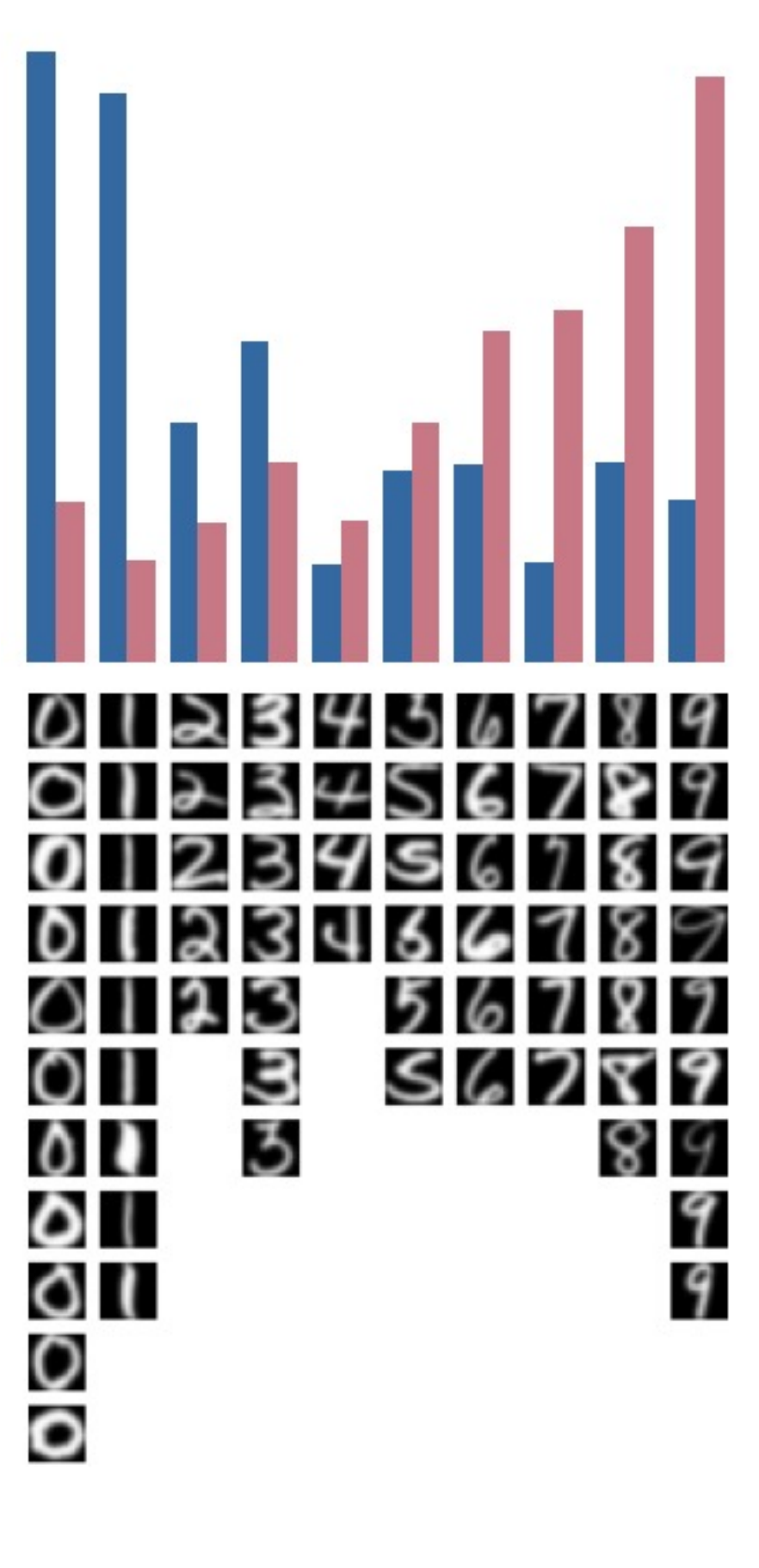}
    \caption{Summary of a dependent MMD coreset for two datasets of handwritten digits. The weights and exemplars from Figure~\ref{fig:usps_hist_1} have been combined based on their label. Weights for $X_a$ are shown in blue {\color{bluetol}\rule{\widthof{block}}{\heightof{block}}} and weights for $X_b$ are shown in red {\color{redtol}\rule{\widthof{block}}{\heightof{block}}}.}
    \label{fig:usps_hist_2}
\end{figure}

We begin by generating a dependent MMD coreset for the two datasets, with $\epsilon^2 = 0.005$. The coreset was selected from the union of $X_b$ and the USPS test set; candidates from $X_a$ were not included as these will later be used to train the algorithms to be assessed. Figure~\ref{fig:usps_hist_1} shows the resulting dependent MMD coreset, with the bars showing the weights $w_{a, i}, w_{b,i}$ associated with the two datasets, and the images below the $x$ axis showing the corresponding images $u_i$. In Figure~\ref{fig:usps_hist_2}, the $u_i$ and $w_i$ have been grouped by number, so that if $y(u)$ is the label of image $u$, the $j$th bar for $\mathbb{Q}_a$ has weight $\sum_{i\in S: y(u_i)=j} w_{a, j}$.

We can see that the coreset has selected points that cover the spread of the overall dataset. However, looking at Figure~\ref{fig:usps_hist_2}, we see that the weights assigned to these exemplars in $\mathbb{Q}_a$ and $\mathbb{Q}_b$ mirror the relative frequencies of each digit in the corresponding datasets $X_a$ and $X_b$ (Figure~\ref{fig:number_hist}).

\vspace{.1in}

We will now make use of this dependent coreset to explore the generalization performance of three classifiers. In this case, since we have labeled data, we can qualitatively assess generalization performance as a benchmark. We trained three classifiers---a decision tree with maximum depth of 8, a random forest with 100 trees, and a multilayer perceptron (MLP) with a single hidden layer with 100 units---on $X_a$ and the corresponding labels. In each case, we used the implementation in \texttt{scikit-learn} \cite{scikit-learn}, with parameters chosen to have similar train set accuracy on $X_a$. Table~\ref{tab:accuracies} shows the associated classification accuracies on the datasets $X_a$ and $X_b$.

\begin{figure}[h]
    \centering
    \includegraphics[width=.7\textwidth]{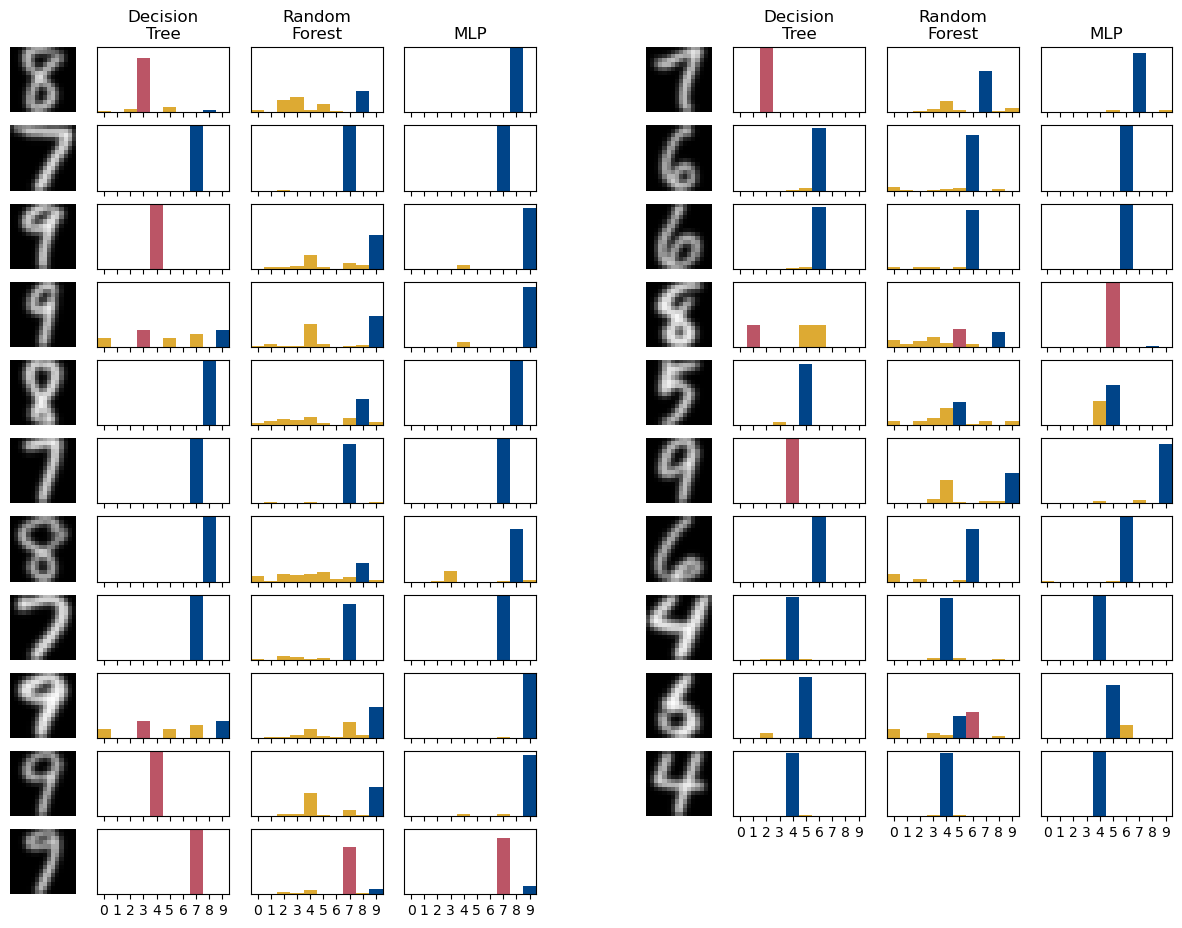}
    \caption{Exemplars over-represented in $\mathbb{Q}_b$, with class probabilities under three algorithms trained on $X_a$. The true class is shown in blue {\color{bluetol}\rule{\widthof{block}}{\heightof{block}}} ; where the highest probability class differs from the true class, the highest probability class is shown in red {\color{redtol}\rule{\widthof{block}}{\heightof{block}}}.}
    \label{fig:overrepresented_exemplars_b}
\end{figure}

\begin{table}[h]
    \centering
    \begin{tabular}{c c c}
        Model & Accuracy on $X_a$ & Accuracy on $X_b$\\
        \hline
        MLP & 0.9998 & 0.8531\\
        Random Forest & 1. & 0.7129 \\
        Decision Tree & 0.9585 & 0.5880
    \end{tabular}
    \caption{Accuracies of three classification algorithms, on datasets $X_a$ and $X_b$}
    \label{tab:accuracies}
\end{table}
\begin{figure}
    \centering
    \includegraphics[width=.7\textwidth]{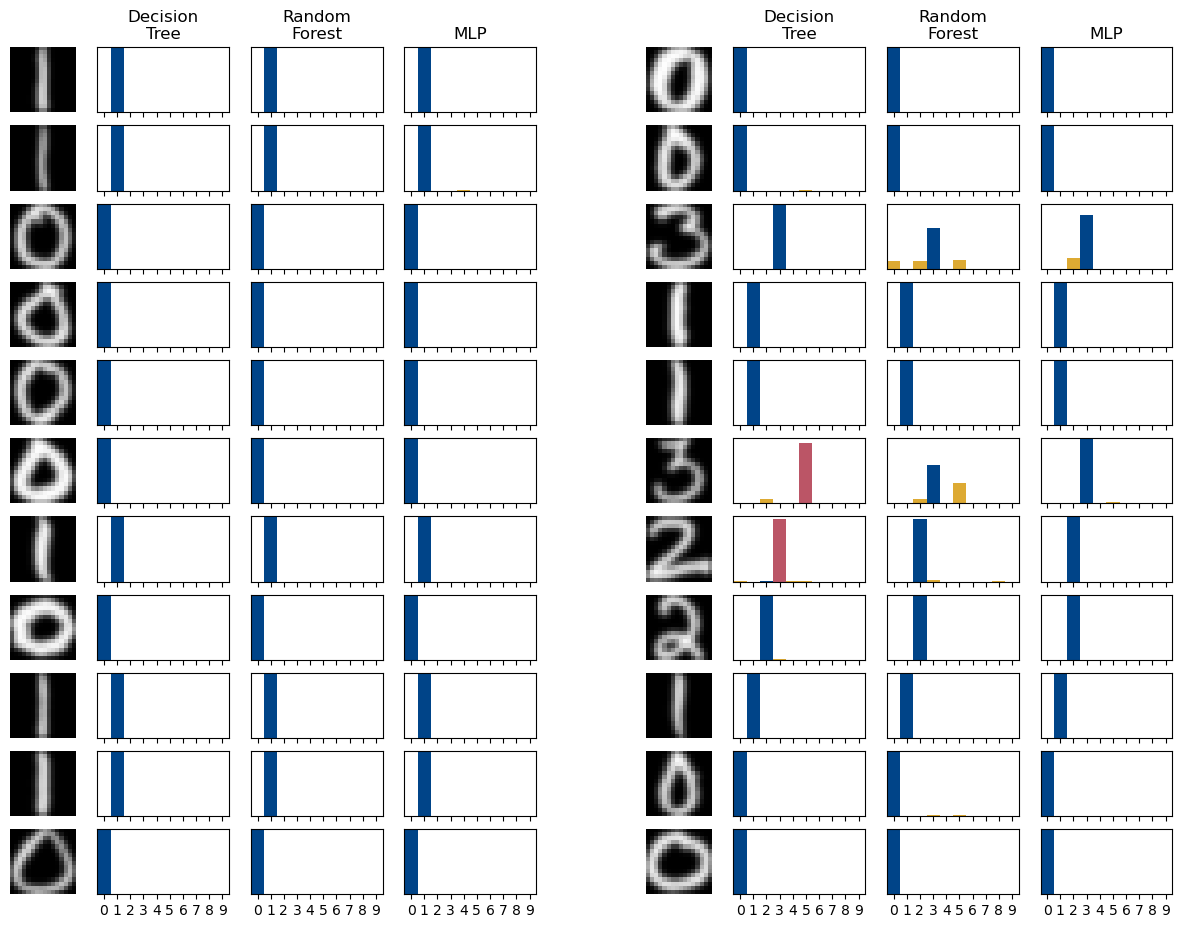}
    \caption{Exemplars over-represented in $\mathbb{Q}_a$, with class probabilities under three algorithms trained on $X_a$. The true class is shown in blue {\color{bluetol}\rule{\widthof{block}}{\heightof{block}}} ; where the highest probability class differs from the true class, the highest probability class is shown in red {\color{redtol}\rule{\widthof{block}}{\heightof{block}}}.}
    \label{fig:overrepresented_exemplars_a}
\end{figure}
We then considered all points $u_i$ in our dependent coreset $(\mathbb{Q}_a, \mathbb{Q}_b)$ where $f_i = w_{b, i} / w_{a, i} > 2$---i.e., points that are much more representative of $X_b$ than $X_a$. We then looked at the class probabilities of the three algorithms, on each of these points, as shown in Figure~\ref{fig:overrepresented_exemplars_b}. We see that the decision tree mis-classifies nine of the 21 exemplars, and is frequently highly confident in its misclassification. The random forest misclassifies three examples, and the MLP misclassifies two. We see this agrees with the ordering provided by empirically evaluating generalization in Table~\ref{tab:accuracies}---the MLP generalizes best, and the decision tree worst. Across all algorithms, we see that performance on the numbers 8 and 9 seems to generalize worst---to be expected, since these digits are most underrepresented in $X_a$.

For comparison, in Figure~\ref{fig:overrepresented_exemplars_a} we show the points where $f_i < 0.5$---i.e., points that are much more representative of $X_a$ than $X_b$. Note that, since our candidate set did not include any members of $X_a$, none of these points were in our training set. Despite this, the accuracy is high, and fairly consistent between the three classifiers (the decision tree misclassifies two exemplars; the other two algorithms make no errors).

\begin{figure}
    \centering
    \includegraphics[width=.7\textwidth]{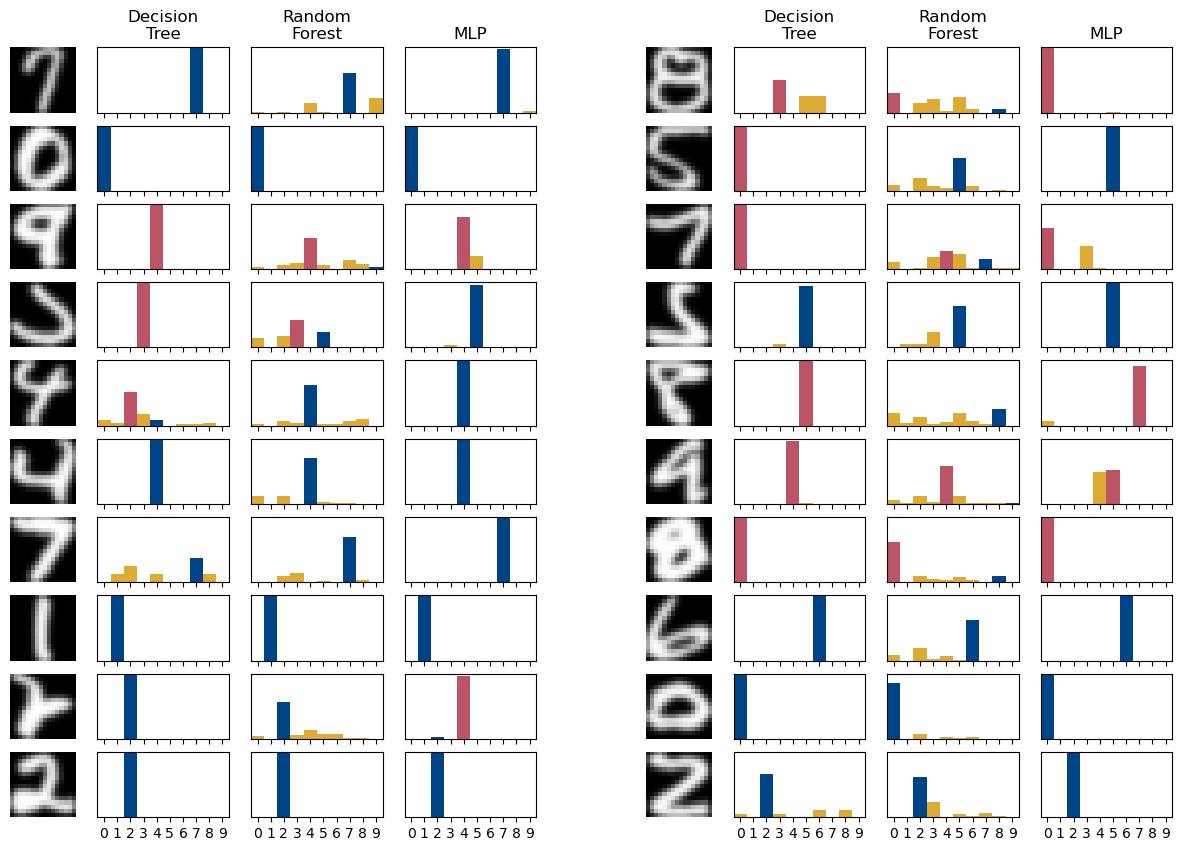}
    \caption{Criticisms of $\mathbb{Q}_b$ from the dataset $X_b$, with class probabilities under three algorithms trained on $X_a$. The true class is shown in blue {\color{bluetol}\rule{\widthof{block}}{\heightof{block}}} ; where the highest probability class differs from the true class, the highest probability class is shown in red {\color{redtol}\rule{\widthof{block}}{\heightof{block}}}.}
    \label{fig:critics_b}
\end{figure}

The dependent coreset only provides information about performance on ``representative'' members of $X_b$. Since classifiers will tend to underperform on outliers, looking only at the dependent MMD coreset does not give us a full picture of the expected performance. 
We can augment our dependent MMD coreset with criticisms---points that are poorly described by the dependent coreset. Figure~\ref{fig:critics_b} shows the performance of the three algorithms on a size-20 set of criticisms for $X_b$. Note that, overall, accuracy is lower than for the coreset---unsurprising, since these are outliers. However, as before, we see that the decision tree performs worst on these criticisms (nine mis-classifications), with the other two algorithms performing slightly better (six mis-classifications for the random forest, and seven for the MLP).

Note that, since accuracy does not correspond to a function in a RKHS, we cannot expect to use the coreset to bound the expected accuracy of an algorithm on the full dataset. Indeed, while the coresets and critics correctly suggest that the decision tree generalizes poorly, they do not give conclusive evidence on the relative generalization abilities of the other two algorithms. However, they do highlight what sort of data points are likely to be poorly modeled under each algorithm. By providing a qualitative assessment of performance modalities and failure modes on either typical points for a dataset, or points that are disproportionately representative of a dataset (vs the original training set) dependent MMD coresets allows users to identify potential generalization concerns for further exploration.

\subsection{Discussion}
MMD coresets have already proven to be a useful tool for summarizing datasets and understanding models. However, as we have shown in Section~\ref{sec:results_interpretable}, their interpretability wanes when used to compare related datasets. Dependent MMD coresets provide a tool to jointly model multiple datasets using a shared set of exemplars, facilitating comparison of the coresets. By considering the relative weights of exemplars across two dataset, we can identify areas of domain mis-match. By exploring performance of algorithms on such points, we can glean insights about the ability of a model to generalize to new datasets.

Dependent MMD coresets are just one example of a dependent coreset that could be constructed using this framework. Future directions include exploring dependent analogues of other measure coresets \cite{ClaSol2018}.


%\externalbibliography{yes}
%\bibliography{your_external_BibTeX_file}

%=====================================
% References, variant B: internal bibliography
%=====================================
\bibliographystyle{plainnat}
\bibliography{coresets}

% If authors have biography, please use the format below
%\section*{Short Biography of Authors}
%\bio
%{\raisebox{-0.35cm}{\includegraphics[width=3.5cm,height=5.3cm,clip,keepaspectratio]{Definitions/author1.pdf}}}
%{\textbf{Firstname Lastname} Biography of first author}
%
%\bio
%{\raisebox{-0.35cm}{\includegraphics[width=3.5cm,height=5.3cm,clip,keepaspectratio]{Definitions/author2.jpg}}}
%{\textbf{Firstname Lastname} Biography of second author}

% The following MDPI journals use author-date citation: Arts, Econometrics, Economies, Genealogy, Humanities, IJFS, JRFM, Laws, Religions, Risks, Social Sciences. For those journals, please follow the formatting guidelines on http://www.mdpi.com/authors/references
% To cite two works by the same author: \citeauthor{ref-journal-1a} (\citeyear{ref-journal-1a}, \citeyear{ref-journal-1b}). This produces: Whittaker (1967, 1975)
% To cite two works by the same author with specific pages: \citeauthor{ref-journal-3a} (\citeyear{ref-journal-3a}, p. 328; \citeyear{ref-journal-3b}, p.475). This produces: Wong (1999, p. 328; 2000, p. 475)

%%%%%%%%%%%%%%%%%%%%%%%%%%%%%%%%%%%%%%%%%%
%% for journal Sci
%\reviewreports{\\
%Reviewer 1 comments and authors’ response\\
%Reviewer 2 comments and authors’ response\\
%Reviewer 3 comments and authors’ response
%}
%%%%%%%%%%%%%%%%%%%%%%%%%%%%%%%%%%%%%%%%%%
\end{document}